\documentclass[showpacs,preprintnumbers,amsmath,amssymb,floatfix]{revtex4}
\usepackage{graphicx}
\usepackage{dcolumn}
\usepackage{bm}
\usepackage{epsfig}

\newcommand{\no}{\nonumber}

\begin{document}
\title{Condensation of Ideal Bose Gas Confined in a Box Within a Canonical Ensemble}

\author{Konstantin Glaum}
\email{glaum@physik.fu-berlin.de}
\affiliation{%
Freie Universit{\"a}t Berlin, Institut f{\"u}r Theoretische Physik,
Arnimallee 14, 14195 Berlin, Germany}%
\author{Hagen Kleinert}
\email{kleinert@physik.fu-berlin.de}
\affiliation{%
Freie Universit{\"a}t Berlin, Institut f{\"u}r Theoretische Physik,
Arnimallee 14, 14195 Berlin, Germany}%
\author{Axel Pelster}
\email{axel.pelster@uni-duisburg-essen.de}
\affiliation{%
Universit{\"a}t Duisburg-Essen, Campus Duisburg, Fachbereich Physik,
Lotharstra{\ss}e 1, 47048 Duisburg, Germany}%

\date{\today}

\begin{abstract}
We set up recursion relations for the partition function and the ground-state
occupancy for a fixed number of non-interacting bosons confined in a square
box potential and determine the temperature dependence of
the specific heat and the particle number in the ground state.
A proper semiclassical treatment is set up which yields
the correct small-$T$-behavior
in contrast to an earlier theory in Feynman's textbook on Statistical
Mechanics, in which the special role of the ground state was ignored.
The results are compared with an exact quantum mechanical treatment. 
Furthermore, we derive the finite-size effect of the system.
\end{abstract}
\maketitle
\section{Introduction}
The thermodynamic properties of a Bose gas are usually studied in the
grand-canonical formulation where energy and particle
number are fixed on the average.
Present-day experiments, on the other hand, are performed in a magnetic or
optical trap which contain a roughly fixed number
of particles. Thus, the experimental situation may be
better approximated by a canonical ensemble with fixed particle number.
In the thermodynamic limit
one usually expects
grand-canonical and canonical
treatments to yield the same results for
all thermodynamic quantities
and the condensate fraction.
This has indeed been proved in \cite{Ziff} in general and for  
harmonic traps in
\cite{Politzer,Gajda,Grossman,Weiss}.
Some statistical quantities, however,
do not have the same thermodynamic limits
in the two ensembles. One fundamental example is
the fluctuation width 
$ \Delta N_0\equiv  \sqrt{\langle  (\Delta N_0)^2\rangle}$
of the particle-number $N_0$ in the ground state. 
In the grand-canonical ensemble, this quantity has the size
$\sqrt{N_0(N_0+1)}$ and is at low temperatures $T$
of the order of the total  particle
number $N$ \cite{Ziff,Politzer,Johnston,terHaar,Yukalov1}.
In the canonical ensemble, on the other hand,
this quantity can be shown to
vanish at $T=0$ since $N_0$ becomes equal to the fixed particle number $N$.
The grand-canonical result is a consequence
of the assumed statistical independence
of all open grand-canonical subsystems.
This assumption is only fulfilled if there is no
long-range ordered condensate, otherwise it is wrong.
The problem can be
properly removed by fixing the whole particle number and ignoring its
fluctuations as was shown e.g. in \cite{Ziff,Johnston}. The particle number
in the ground state is then $N_0 = N - N_{\rm ex}$, where $N_{\rm ex}$ is the
excited particle number. Since, furthermore, the fluctuation 
$\Delta N_{\rm ex}$ of the thermal gas, consisting of excited 
states, can be shown to vanish for zero temperature, the same is true for
$\Delta N_0$. Its temperature behavior is found to be
$\Delta N_0 \sim (T/T_c)^{3/2} \,N^{1/2}$
for an ideal gas in a harmonic trap \cite{Politzer} and
$\Delta N_0 \sim (T/T_c) \, N^{2/3}$ for a homogeneous system
\cite{Hauge}
for temperatures $T$ below its critical value $T_c$.
By fixing the full particle number $N$, the system
does not exchange its particles with the environment at all and this situation, therefore,
corresponds to one in a canonical ensemble. The only
exception is that excited states are then still described by the 
grand-canonical Bose-Einstein distribution. The entirely canonical 
description will be presented below.

The canonical treatment is conveniently performed in a first-quantized
path-integral description. All particles are represented
by cycles winding around a cylinder in Euclidean spacetime
whose circumference is the
imaginary-time axis $\tau \in (0,\hbar /k_BT) $ \cite{Feynman,Kleinert1}. For sufficiently low
temperatures,
the indistinguishability
of particles leads to
long closed paths winding many times around this cylinder,
giving rise to correlated subsets of particles
\cite{Ceperley}. In Ref.~\cite{Bund}, this representation was
utilized to claim that condensation amounts directly to a proliferation
of long cycles. This claim can, however, not be upheld.
The formation of correlated collectives is the origin
of superfluidity,
not condensation, as it was correctly implemented
for an ideal gas in a harmonic trap \cite{Schneider}.
The essential signal for the Bose-Einstein condensation is, instead,
associated with the occupancy of the single-particle ground state.
Thus, not the winding numbers of the cycles are relevant to condensation,
but the weight of the ground state in them. For a harmonic trap, this
was
properly shown in Refs.~\cite{Landsberg,Wilkens}.

If the harmonic trap is replaced by a square box,
whose infinite-volume limit is the homogeneous system,
the systems confined in it show some unusual features. 
It is the purpose of this paper to exhibit these.
The discussion starts in Section \ref{PROP} with a brief
summary of the imaginary-time evolution amplitude for a fixed number
$N$ of bosons, where the indistinguishability of particles is explicitly
taken into account. In Section \ref{CYCLES} we calculate the partition
function of such a canonical system from the trace of the $N$-particle
imaginary-time evolution amplitude
written in the cycle representation of paths.
Since this
is hard to evaluate numerically for a large particle number, we have
derived an
efficient recursion relation along the lines of 
Refs.~\cite{Weiss,Sato,Borrmann,Devreese1,Vorontsov}
in Section \ref{GRAND}.
In Section \ref{SPECTR} this relation is made specific by
a spectral decomposition of the canonical partition function. This
allows us in Section \ref{KONDENS} to extract the ground-state contribution
to the partition function from the cycle representation, from which we read
off the probability to find a particle in the ground state.
This is used to define
the condensate fraction $N_0/N$ for a system with a fixed particle number $N$.
In Section \ref{POT} we apply our results to the Bose gas confined
in the square-box potential.
There
we show that the semiclassical approximation requires
a delicate treatment of the ground state for finite systems.
This is necessary in order to obtain the correct low-temperature behavior of
the thermodynamical functions in contrast to an earlier theory in Feynman's 
original textbook
\cite{Feynman}.
The results are then compared with an exact quantum mechanical calculation.
In the subsequent Section \ref{Tkrit} we investigate the finite-size effect
in a box potential. We identify a would-be transition temperature
as the finite-$N$ version of the critical temperature
of Bose-Einstein condensation in a canonical ensemble.
Subsequently, canonical results calculated for these crossover temperatures are
compared with grand-canonical ones.
We end with  an analytic calculation
of the finite-size effect upon the critical temperature in the grand-canonical
ensemble.
\section{Quantum Statistics of Identical Particles}\label{PROP}
We start by briefly reviewing the quantum statistics of a fixed number
$N$ of non-interacting identical particles in a first quantized approach
\cite{Feynman,Kleinert1}.
At first, they are treated
as distinguishable particles. Their orbits in imaginary time $\tau$
are denoted by ${\bf x}_\nu (\tau)$ with the particle indices
$\nu=1,2,\ldots,N$.
Global and local thermodynamic properties of the many-particle system
are determined by the imaginary-time evolution amplitude which is given by
the path integral
\begin{eqnarray}
\label{PID}
({\bf x}_{1b} , \ldots ,{\bf x}_{Nb},\tau_b | {\bf x}_{1a} , \ldots ,
{\bf x}_{Na},\tau_a ) = \left\{ \prod_{\nu=1}^N \int_{{\bf x}_\nu (\tau_a)
= {\bf x}_{\nu a}}^{{\bf x}_\nu (\tau_b) = {\bf x}_{\nu b}}
{\cal D} {\bf x}_\nu \right\} \, e^{- {\cal A}[{\bf x}_1, \ldots ,
{\bf x}_N]/ \hbar} \, .
\end{eqnarray}
The Euclidean action in the exponent has the generic form
\begin{eqnarray}
\label{IA}
{\cal A}[{\bf x}_1, \ldots , {\bf x}_N ] = \sum_{\nu=1}^N
\int_{\tau_a}^{\tau_b} d \tau \left[ \frac{M}{2} \dot{\bf x}^2_\nu ( \tau )
+ V \left( {\bf x}_\nu(\tau)\right) \right]
\end{eqnarray}
with the particle mass $M$ and the common background potential
$V \left( {\bf x}_\nu \right)$ for all particles.
Due to the additivity of the action (\ref{IA}) we find that the
imaginary-time evolution amplitude of $N$ distinguishable non-interacting
particles factorizes into $N$ one-particle amplitudes
\begin{eqnarray}
\label{QS9}
({\bf x}_{1b} , \ldots ,{\bf x}_{Nb},\tau_b | {\bf x}_{1a} , \ldots ,
{\bf x}_{Na},\tau_a )
=({\bf x}_{1b} , \tau_b | {\bf x}_{1a} , \tau_a ) \, \cdots \,
({\bf x}_{Nb} , \tau_b | {\bf x}_{Na} , \tau_a ) \, ,
\end{eqnarray}
These one-particle amplitudes fulfill the semi-group property
\begin{eqnarray}
\label{QS7}
({\bf x}_c , \tau_c | {\bf x}_a , \tau_a ) = \int d^3 x_b \,
({\bf x}_c , \tau_c | {\bf x}_b , \tau_b )\,({\bf x}_b , \tau_b | {\bf x}_a ,
\tau_a )
\end{eqnarray}
and are invariant with respect to translations in the
imaginary time:
\begin{eqnarray}
\label{QS3}
({\bf x}_b , \tau_b+\tau_0 | {\bf x}_a , \tau_a+\tau_0 ) =
({\bf x}_b , \tau_b | {\bf x}_a , \tau_a ) \, .
\end{eqnarray}
We now impose the indistinguishability of particles.
In $D>2$ space dimensions there are only
two kinds of indistinguishable particles: bosons with a completely symmetric
and fermions with a completely antisymmetric $N$-particle
wave function. According to Pauli's spin-statistic-theorem they are
associated with integer
and half-integer spins, respectively. The corresponding modifications
of the path integral (\ref{PID}) are straight-forward. For bosons, we have
to sum over all permuted final positions
${\bf x}_{P(\nu) b}$, where $P(\nu)$ denotes one of the $N!$ possible
permutations of the index $\nu$.
For fermions, there is an extra minus sign for odd permutations.
Restricting ourselves in this paper to the bosonic case, we must calculate
the imaginary-time evolution amplitude
\begin{eqnarray}
\label{PIID}
({\bf x}_{1b} , \ldots ,{\bf x}_{Nb},\tau_b | {\bf x}_{1a} , \ldots ,
{\bf x}_{Na},\tau_a )^B = \, \frac{1}{N!} \sum_P 
({\bf x}_{P(1)b} , \ldots ,{\bf x}_{P(N)b},\tau_b | {\bf x}_{1a} , \ldots ,
{\bf x}_{Na},\tau_a ) \, .
\end{eqnarray}
%
\section{Cycle Expansion}\label{CYCLES}
\begin{figure}[t]
$\mbox{}$\\[1cm]
\begin{center}
\begin{picture}(350,100)
\includegraphics[scale=0.9]{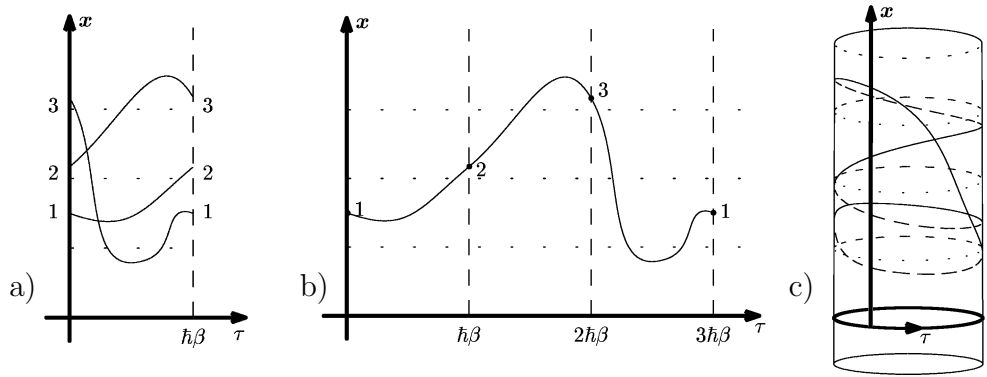}
\put(-370,30){\large a)}
\put(-260,30){\large b)}
\put(-75,30){\large c)}
\end{picture}
\end{center}
\caption{\label{PFAD} Example of paths contributing to (\ref{QS19}) for
a cycle of length $n=3$: a)
final points coincide with initial points of the other particles,
b) and c) show the same situation in an extended zone
scheme and wrapped upon a cylinder, respectively. Figure after 
Ref.~\cite{Bund}.}
\end{figure}

The canonical partition function of $N$
non-interacting bosons is given by the trace
\begin{eqnarray}
\label{QS17}
Z_N^B (\beta)= \int d^3 x_1 \, \cdots \, \int d^3 x_N \, ({\bf x}_{1} , \ldots
,{\bf x}_{N},\hbar \beta | {\bf x}_{1} , \ldots , {\bf x}_{N},0 )^B  \, ,
\end{eqnarray}
where $\beta \equiv (k_B T)^{-1}$ represents the reciprocal temperature.
The calculation of this basic statistic quantity is non-trivial
due to the indistinguishability of the particles.
At first we consider the Bose gas with small particle numbers $N$.
For $N=1$, the partition function (\ref{QS17}) reduces to the well-known
expression
\begin{eqnarray}
\label{QS16}
Z_1 ( \beta ) = \int d^3 x \,({\bf x} ,\hbar \beta | {\bf x} , 0 ) \, .
\end{eqnarray}
Here we have omitted the index $B$, as the distinguishability of particles
is not an
issue for a single particle.
For $N=2$, we obtain from (\ref{QS17}) according to the indistinguishability
(\ref{PIID}) and the factorization property (\ref{QS9})
\begin{eqnarray}
Z_2^{B}(\beta) = \frac{1}{2!} \, \int d^3 x_1 d^3 x_2 \Big\{
({\bf x}_{1} ,\hbar \beta | {\bf x}_{1} , 0 ) \,
({\bf x}_{2} ,\hbar \beta | {\bf x}_{2} , 0 )
+ \, ({\bf x}_{2} ,\hbar \beta | {\bf x}_{1} , 0 ) \,
({\bf x}_{1} ,\hbar \beta | {\bf x}_{2} , 0 ) \Big\} \, .
\label{QS14}
\end{eqnarray}
Using the translational invariance (\ref{QS3}) together with the semi-group
property (\ref{QS7}) for the second term, this yields with (\ref{QS16})
\begin{eqnarray}
\label{QS162}
Z_2^B(\beta) = \frac{1}{2} \Big\{ Z_1^2 ( \beta ) + \, Z_1 (2 \beta )
\Big\} \,.
\end{eqnarray}
In a similar way, we obtain the result for $N=3$:
\begin{eqnarray}
\label{QS16E}
Z_3^B  (\beta)= \frac{1}{6} \Big\{ Z_1^3 ( \beta ) + 3  \,
Z_1 (\beta )\,Z_1 (2 \beta )
+ 2  Z_1 ( 3 \beta ) \Big\} \,.
\end{eqnarray}
For an arbitrary number $N$ of bosons we must calculate according to
(\ref{QS9}) and (\ref{PIID}) the
canonical partition function (\ref{QS17})
\begin{eqnarray}
Z_N^B (\beta) = \frac{1}{N!} \, \sum_{P} 
\int d^3 x_1 \, \cdots \, \int d^3 x_N \,
({\bf x}_{P(1)} , \hbar \beta | {\bf x}_{1} , 0 ) \, \cdots \,
({\bf x}_{P(N)} , \hbar \beta | {\bf x}_{N} , 0 ) \, .
\label{QS18}
\end{eqnarray}
Due to the indistinguishability of the particles there are single-particle
amplitudes contributing to the partition function (\ref{QS18}), whose final
points coincide with the initial points of other particles.
These amplitudes can be combined to many-particle formations, which can be
represented by paths winding around a cylinder of circumference $\hbar \beta$.
A typical situation for three particles is shown in Figure \ref{PFAD}.
This represents a single closed cycle with three windings around the
cylinder.
An arbitrary single cycle with $n$ windings
is given by the multiple integral
\begin{eqnarray}
Z_n^{\rm c}(\beta) = \int d^3 x_1 \, \cdots \, \int d^3 x_n \,
({\bf x}_1 , \hbar \beta | {\bf x}_n , 0 ) \, ({\bf x}_n , \hbar \beta |
{\bf x}_{n-1} , 0 ) \, \cdots \, ({\bf x}_3 , \hbar \beta | {\bf x}_2 , 0 )
\, ({\bf x}_2 , \hbar \beta | {\bf x}_1 , 0 ) \, .
\label{QS19}
\end{eqnarray}
Due to the translational invariance (\ref{QS3}) and the semi-group property
(\ref{QS7}) of the 1-particle imaginary-time
evolution amplitude we see that
\begin{eqnarray}
\label{QS21}
Z_n^{\rm c}(\beta) =  Z_1 (n \beta ) \, .
\end{eqnarray}
Thus, a contribution of a closed cycle of length $n$
corresponds to a simple one-particle partition function but with a
temperature scaled down by the same factor $n$. With this we return
to the general expression (\ref{QS18}) for the canonical partition function.
The full $N$-particle partition function decomposes
into mutually disconnected groups of closed paths
winding around a cylinder of circumference $\hbar \beta$. Due to the
indistinguishability of particles this decomposition is nontrivial since
the cycles may have different winding numbers.

\begin{figure}[t]
\begin{center}
\includegraphics[scale=0.7]{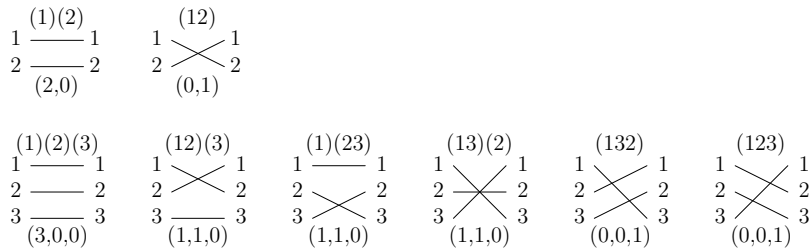}
\end{center}
\caption{\label{N23}Permutations for $N=2$ (first row) and $N=3$ (second row)
particles. The indices above each permutation stand for the cycle notation,
whereas the respective $N$-tupel $(C_1, \ldots , C_N)$ below each permutation
denotes the cycle number notation.}
\end{figure}

In order to illustrate the situation, we consider in detail the possible path
combinations for $N=2$ and $N=3$ particles. Figure \ref{N23} shows the
respective permutations in different notations.
In the cycle notation the indices of particles contributing to the closed
paths are stated within parentheses. This cycle notation is
mathematically useful but is not adopted to the indistinguishability of
particles as it relies explicitly on the numbering of the particles.
For instance, the second, third, and forth
permutation in the second row of Figure \ref{N23} are regarded as different
permutations, although they are identical from the physical point of view.
This motivates to introduce a different notation for the
permutations which respects
the indistinguishability of the particles. It is the cycle-number
notation which characterizes each permutation by the $N$-tupel
$(C_1, \ldots , C_N)$ of numbers $C_n$ of cycles with the length $n$.
In this notation the second, third, and forth
permutations in the second row of Figure \ref{N23} have the same 3-tupel
$(1,1,0)$.\\
Using the result (\ref{QS21}) for the multiple integral (\ref{QS19})
we arrive at the cycle decomposition of
the canonical partition function (\ref{QS18}):
\begin{eqnarray}
\label{QS22}
Z_N^B (\beta) = \frac{1}{N!} \, \sum_P 
\prod_{n=1}^{\Sigma C_n n = N} \, \left[
Z_1(n \beta)\right]^{C_n} \, .
\end{eqnarray}
Note that the numbers of cycles $C_n$ of length $n$, which occur for
a certain permutation $P$, are restricted by the condition
$\Sigma_n C_n n = N$.
The sum in (\ref{QS22}) can be rearranged as follows.
Instead of summing over permutations and decomposing them into cycles, we
may sum directly
over all different cycle numbers $(C_1, \ldots , C_N)$ which respect the
equation $\Sigma_n C_n n = N$.
By doing so, we have to take into account that a particular
configuration $(C_1, \ldots , C_N)$ can occur with a certain multiplicity
$M(C_1,\ldots,C_N)$. For instance, in Figure \ref{N23} we can read off
$M(3,0,0)=1$, $M(1,1,0)=3$, and $M(0,0,1)=2$ for three particles. Thus, we
obtain
\begin{eqnarray}
\label{QS23}
Z_N^B (\beta) = \frac{1}{N!} \, \sum_{C_1,\ldots,C_N}^{\Sigma C_n n = N}
M(C_1,\ldots,C_N)\, \prod_{n=1}^{\infty} \, \left[Z_1(n \beta)\right]^{C_n}
\, ,
\end{eqnarray}
where we have formally extended the product to all integers by setting $C_n=0$
for any $n > N$ due to the condition $\Sigma_n C_n n = N$.
Now we derive a general formula for the multiplicities $M(C_1,\ldots,C_N)$.
In total the
cycle numbers $(C_1,\ldots,C_N)$ can be arranged in $N!$ different
combinations. However, not all of those represent different permutations
as we shall demonstrate by the
following examples:
\begin{itemize}
\item The cycle numbers $(C_1=0, C_2 = 0,C_3=1)$ allow for 6 combinations
$(123)$, $(312)$, $(231)$, $(132)$, $(213)$, $(321)$, where the former and
latter three combinations are different representations of only two
permutations
\begin{eqnarray}
P_1 = \left( \,\begin{array}{@{}ccc} 1 & 2 & 3 \\ 3 & 1 & 2
\end{array} \right) \, , \hspace*{1cm}
P_2 = \left( \, \begin{array}{@{}ccc} 1 & 2 & 3 \\ 2 & 3 & 1 \end{array}
\right) \, .
\end{eqnarray}
\item The cycle numbers $(C_1=3,C_2=0,C_3=0)$ lead to 6 combinations
$(1)(2)(3)$, $(1)(3)(2)$, $(2)(1)(3)$, $(2)(3)(1)$, $(3)(1)(2)$, $(3)(2)(1)$
which are all representations of the same permutation
\begin{eqnarray}
P_3 = \left( \,\begin{array}{@{}ccc} 1 & 2 & 3 \\ 1 & 2 & 3 \end{array}
\right) \, .
\end{eqnarray}
\end{itemize}
These observations are generalized as follows:
\begin{itemize}
\item A cyclic permutation within a $n$-cycle does not lead to a new
permutation, so we have $n$ irrelevant combinations.
\item Exchanging cycles of the same length does not lead to a new permutation,
so we have $C_n!$ irrelevant combinations.
\end{itemize}
Summarizing these combinatorial considerations, we obtain for the multiplicity
the Cauchy formula of the permutation group \cite{Cauchy1,Cauchy2}:
\begin{eqnarray}
\label{MULT1}
M(C_1,\ldots,C_N) = \frac{N!}{\prod_n C_n!\,n^{C_n}} \, .
\end{eqnarray}
After this combinatorial excursion we return to the canonical partition
function (\ref{QS23}).
Using  (\ref{MULT1}), we find our second main result in
form of a cycle representation of the canonical partition function
\cite{Feynman}:
\begin{eqnarray}
\label{QS27}
Z_N^B (\beta) = \sum_{C_1,\ldots,C_N}^{\Sigma C_n n = N}
\prod_{n=1}^{\infty} \,\frac{1}{C_n!}\,
\left[ 
\, \frac{Z_1(n \beta)}{n} \, \right]^{C_n} \, .
\end{eqnarray}
Note that specializing (\ref{QS27}) for $N=2$ and $N=3$ particles reproduces,
indeed, the earlier findings (\ref{QS162}) and (\ref{QS16E}).

Although the cycle representation allows us, in principle, to
determine the canonical partition function $Z_N^B(\beta)$, this
approach is not efficient for a large particle number $N$ as the computational
effort grows factorially with $N$.
Furthermore, it is rather
cumbersome to figure out all cycle numbers $C_n$ which
respect the condition  $\Sigma_n C_n n = N$. Therefore, we need to find a
more efficient algorithm to determine canonical partition functions. This
is done by using the fact that the grand-canonical partition function is the
generating function of all canonical partition functions \cite{Devreese1}.
\section{Generating Function and Recursion Relation}\label{GRAND}
Introducing the fugacity $z \equiv e^{\beta \mu}$ with the chemical potential
$\mu$, the grand-canonical partition function reads
\begin{eqnarray}
\label{REK1}
{\cal Z}_{GC}^B ( \beta , z ) = \sum_{N=0}^\infty Z_N^B(\beta) \, z^N\, .
\end{eqnarray}
This quantity serves as the generating function of canonical
partition functions $Z_N^B(\beta)$, which are simply
the Taylor expansion coefficients
\begin{eqnarray}
\label{REK2}
Z_N^B (\beta) = \frac{1}{N!}\, \left. \frac{\partial^N
{\cal Z}_{GC}^B ( \beta , z )}{\partial z^N} \right|_{z=0}\, .
\end{eqnarray}
In order to evaluate (\ref{REK1}), we use the cycle representation
(\ref{QS27}) and
insert the condition $N=\Sigma C_n n$ into the exponent of the fugacity $z$.
Thus, we obtain at first
\begin{eqnarray}
\label{REK3}
{\cal Z}_{GC}^B ( \beta , z ) = \sum_{N=0}^{\infty}
\sum_{C_1,\ldots,C_N}^{\Sigma n C_n = N}  \prod_{n=1}^{\infty}
\frac{1}{C_n!} \, \left[ 
\, \frac{Z_1(n \beta)z^n}{n} \right]^{C_n} = \sum_{C_1=0}^{\infty}
\sum_{C_2=0}^{\infty} ...  \prod_{n=1}^{\infty}
\frac{1}{C_n!} \, \left[ 
\, \frac{Z_1(n \beta)z^n}{n} \right]^{C_n} \, .
\end{eqnarray}
In the second expression
the condition $\Sigma n C_n = N$ no longer occurs.
Due to this simplification we can rewrite the multiple sum over the cycle
numbers $C_n$ as a product over the individual sums and obtain
\begin{eqnarray}
\label{REK4}
{\cal Z}_{GC}^B ( \beta , z ) =\prod_{n=1}^{\infty} \sum_{C_n=0}^{\infty}
\frac{1}{C_n!} \, \left[ 
\, \frac{Z_1(n \beta)z^n}{n} \right]^{C_n} \, .
\end{eqnarray}
Performing the $C_n$-summation yields the result
\begin{eqnarray}
\label{REK5}
{\cal Z}_{GC}^B ( \beta , z ) = \exp \left\{ \sum_{n=1}^{\infty}
\frac{Z_1(n \beta)z^n}{n} \right\} \, .
\end{eqnarray}
The partition function of the vacuum, where no particle is present ($N=0$),
is obviously found from (\ref{REK2}) with (\ref{REK5}):
\begin{eqnarray}
\label{REK6}
Z_0^B (\beta) = 1 \, .
\end{eqnarray}
For $N \geq 1$, we observe
that the grand-canonical partition function (\ref{REK5}) obeys the
differential equation
\begin{eqnarray}
\label{REK7}
\frac{\partial {\cal Z}_{GC}^B ( \beta , z ) }{\partial z} =
{\cal Z}_{GC}^B( \beta,z ) \, \sum_{k=1}^{\infty}
Z_1(k \beta) \,z^{k-1} \, .
\end{eqnarray}
This is a self-consistent equation for the generating function which we solve
recursively.
Forming the $N$th derivative of (\ref{REK1}) via the
Leibniz rule of differentiation:
\begin{eqnarray}
\label{REK8}
\left. \frac{\partial^N {\cal Z}_{GC}^B ( \beta , z ) }{\partial z^N}
\right|_{z=0} = \sum_{n=1}^N \frac{(N-1)!}{(n-1)!\,(N-n)!} \,
\left. \frac{\partial^{N-n} {\cal Z}_{GC}^B ( \beta , z ) }{\partial z^{N-n}}
\right|_{z=0} \, \sum_{k=1}^{\infty} 
Z_1(k \beta) \, \left. \frac{\partial^{n-1}z^{k-1}}{\partial z^{n-1}}
\right|_{z=0} \, .
\end{eqnarray}
and taking into account the identity
\begin{eqnarray}
\label{REK9}
\left. \frac{\partial^{n-1}z^{k-1}}{\partial z^{n-1}}  \right|_{z=0} =
\, (n-1)! \, \delta_{k,n} \, ,
\end{eqnarray}
we obtain from (\ref{REK2})
the desired
recursion relation for the canonical partition functions $Z_N^B(\beta)$
\cite{Weiss,Sato,Borrmann,Devreese1,Vorontsov}:
\begin{eqnarray}
\label{REK12}
Z_N^B (\beta) = \frac{1}{N} \, \sum_{n=1}^N 
Z_1 (n \beta)\,Z_{N-n}^B(\beta) \, .
\end{eqnarray}
Inserting the starting value (\ref{REK6}), we immediately
reproduce for $N=2$ and $N=3$ the earlier results (\ref{QS162}) and
(\ref{QS16E}). For large particle numbers it can be shown that the
computational effort grows only algebraically with $N$. Thus, iterating
Eq.~(\ref{REK12}) can be done very efficiently, as is further worked out in
Ref.~\cite{Borrmann}.

Once we have obtained the $N$-particle partition function, we can proceed
to calculate all thermodynamic quantities of interest. For instance, in
Section \ref{POT} we shall determine the specific heat capacity according to
\begin{eqnarray}
C_N^B(T) = k_B T \, \frac{\partial^2}{\partial T^2} \left[ T \ln Z_N^B
\right] \, .
\label{C_N}
\end{eqnarray}
\section{Spectral Decomposition}\label{SPECTR}
In order to apply the recursion relation (\ref{REK12})
to specific systems we must calculate the 1-particle partition
function $Z_1(\beta)$ of Eq.~(\ref{QS16}). For this we solve the
time-independent Schr{\"o}dinger equation and use the energy eigenvalues
$E_{\bf k}$ to calculate $Z_1(\beta)$ as a
spectral sum
\begin{eqnarray}
\label{ZDECO}
Z_1 (\beta) = \sum_{\bf k} e^{- \beta E_{\bf k}} \, ,
\label{spectral}
\end{eqnarray}
where ${\bf k}$ stands for the quantum numbers.
The $n$-fold cycle (\ref{QS21})
\begin{eqnarray}
Z_n^{\rm c} (\beta) = \sum_{\bf k} e^{- \beta \left( n E_{\bf k} \right)}
\label{Z1-spectral}
\end{eqnarray}
may be interpreted physically as a partition function (\ref{spectral}) of a
single quasi-particle consisting of a
correlated union of $n$
fundamental particles with the total energy $n E_{\bf k}$.

The spectral representation (\ref{spectral}) leads
via (\ref{REK5}) to the grand-canonical partition function
\begin{eqnarray}
\label{ZGK}
{\cal Z}_{GC}^B (\beta,\mu) = \exp \left\{ - \sum_{\bf k} \ln \Big[ 1 -
e^{-\beta (E_{\bf k}-\mu)} \Big] \right\} \, .
\end{eqnarray}
From it we may derive the average particle number in a
grand-canonical ensemble according to
\begin{eqnarray}
\label{MT6B}
\langle N \rangle = \frac{1}{\beta} \, \frac{\partial}{\partial \mu}
\left[ \ln {\cal Z}_{GC}^B(\beta,\mu) \right] =
\sum_{\bf k} \frac{1}{e^{\beta(E_{\bf k}-\mu)} - 1} \,\, ,
\end{eqnarray}
which is the well-known
Bose-Einstein distribution.
\section{Ground-State Occupancy}\label{KONDENS}
The most important physical quantity in Bose-Einstein condensation is
the condensate fraction $N_0$ which is also called the ground-state
occupancy. In the usual grand-canonical treatment this quantity is found
by solving numerically the Bose-Einstein distribution (\ref{MT6B})
\cite{Devreese1,Vorontsov}. To this end one calculates the chemical
potential $\mu$ for a given temperature $T$ at a fixed average particle
number $\langle N \rangle$.
Given $\mu(T)$ one finds the average number of particles in ground-state
fraction from the equation
$\langle N_0 \rangle = \left[ e^{\beta(E_G - \mu)} - 1 \right]^{-1}$, where
$E_G$ is the ground-state energy.

In the present paper we shall treat this problem canonically at a fixed
particle number $N$, without fixing a
chemical potential. To this end we calculate directly
the probability that one of the bosons resides in
its ground state \cite{Landsberg,Wilkens}. This probability multiplied by $N$
will be considered to be the finite-$N$ version of the condensate fraction
$N_0$.
In order to determine this quantity, we isolate in
Section \ref{PROBABILITY} the ground-state contribution to the
cycle representation (\ref{QS27}).
Subsequently, in Section \ref{OCCUPANCY}, we derive a more efficient
equation for the canonical ground-state occupancy.
\subsection{Cycle Representation}\label{PROBABILITY}
Using the spectral decomposition (\ref{ZDECO}), we can
rewrite $Z_1(n \beta)$ according to
\begin{eqnarray}
Z_1 (n \beta) = \gamma_n (\beta) + \xi_n (\beta) \, .
\label{h_ntren}
\end{eqnarray}
The first term denotes the contribution of the ground state
\begin{eqnarray}
\gamma_n (\beta) = e^{-n \beta E_{\rm G}} ,
\label{h_ngrund}
\end{eqnarray}
where $E_{\rm G}$ denotes the ground-state energy. The second term in
(\ref{h_ntren}) stems from all excited states
\begin{eqnarray}
\xi_n (\beta) =  \sum_{\bf n} {}^{'} \hspace*{2mm} e^{-n \beta E_{\bf n}} \, ,
\label{h_nrest}
\end{eqnarray}
where the prime indicates the omission of the ground state.
Using the binomial formula, the canonical partition function (\ref{QS27})
is then decomposed as follows:
\begin{eqnarray}
Z_N^B(\beta)  = \sum_{C_1,...C_N}^{\Sigma n C_n =N} \sum_{m_1=0}^{C_1}
... \sum_{m_N=0}^{C_N} \gamma_1^{ \sum n m_n }(\beta) \prod_{n=1}^{N}
\frac{ 
\xi_n^{C_n \!-\! m_n}(\beta) }{ m_n! \, (C_n - m_n)! \, n^{C_n} }  \, ,
\label{Z_Ngrund}
\end{eqnarray}
where we have used the factorization property of the ground-state contribution
(\ref{h_ngrund}), which reads $\gamma_n (\beta) = \gamma_1^n (\beta)$.

Now we are able to determine the weight of the particles
residing in the ground state.
Consider first the cases of $N=2$ and $N=3$ particles, where (\ref{Z_Ngrund})
yields
\begin{eqnarray}
\label{Z_2tren}
Z^B_2 &=& \gamma_1^0 \, \frac{1}{2} \, \Big[ \xi_1^2  +
\, \xi_2 \, \Big] + \gamma_1^1 \, \xi_1 + \gamma_1^2 \, , \\
Z^B_3 &=& \gamma_1^0 \Big[ \, \frac{1}{6} \, \xi_1^3
+ \frac{1}{2} \, \xi_1 \, \xi_2 + \frac{1}{3} \, \xi_3 \, \Big] +
\gamma_1^1 \, \frac{1}{2}\, \Big[  \xi_1^2 + \, \xi_2 \,
\Big] +  \gamma_1^2 \, \xi_1 + \gamma_1^3  \, .
\label{Z_3tren}
\end{eqnarray}
This can, of course, be directly obtained from (\ref{QS162}) and (\ref{QS16E})
by inserting the decomposition (\ref{h_ntren}).
The number of particles in the ground state coincides always with the power of
$\gamma_1$. Thus, the ground-state weights $W_2^B$
and $W_3^B$ associated with $Z_2^B$ and $Z_3^B$ are
\begin{eqnarray}
\label{W_2}
W^B_2  &=&0 \times \gamma_1^0 \, \frac{1}{2} \, \Big[  \xi_1^2 +
\, \xi_2 \Big] + 1 \times \gamma_1^1 \, \xi_1+ 2 \times \gamma_1^2 \, ,\\
W^B_3  &=& 0 \times \gamma_1^0 \, \Big[ \, \frac{1}{6} \, \xi_1^3
+ \frac{1}{2} \, \xi_1 \, \xi_2 + \frac{1}{3} \,\xi_3 \Big] +
1 \times \gamma_1^1 \, \frac{1}{2} \, \Big[  \xi_1^2 + \,
\xi_2 \Big] + 2 \times \gamma_1^2 \, \xi_1 \, + 3 \times
\gamma_1^3  \, .
\label{W_3}
\end{eqnarray}
The general formula for
the weight of the particles in the ground state
is therefore given by
\begin{eqnarray}
W_N^B (\beta)= \sum_{C_1,...C_N}^{\Sigma n C_n =N} \sum_{m_1=0}^{C_1}
... \sum_{m_N=0}^{C_N} \left( \sum_{n=1}^N n m_n \right) \left[
\gamma_1(\beta) \right]^{ \sum n m_n } \prod_{n=1}^{N}
\frac{ 
\xi_n^{C_n \!-\! m_n}(\beta) } { m_n! \, (C_n \!-\! m_n)! \, n^{C_n} } \, .
\label{W_Ngrund}
\end{eqnarray}
\subsection{Recursion Relation}\label{OCCUPANCY}
The evaluation of the cycle representation
(\ref{W_Ngrund}) of the ground-state weight is quite cumbersome for large
particle numbers $N$.
Thus, we aim at deriving a more efficient relation for determining
$W_N^B(\beta)$. Therefore, we observe
that the canonical partition function (\ref{Z_Ngrund})
and the ground-state weight (\ref{W_Ngrund}) are related via
\begin{eqnarray}
W_N^B (\beta) = \gamma_1(\beta) \,\frac{ \partial }{ \partial
\gamma_1(\beta) }   Z_N^B (\beta)\, .
\label{W_N-Z_N}
\end{eqnarray}
It is useful to define a generating function for the ground-state weight
as the sum
\begin{eqnarray}
{\cal W}_{GC}^B (\beta,z) = \sum_{N=0}^{\infty} W_N^B (\beta) \, z^N \, ,
\label{WGK[z]}
\end{eqnarray}
from which we conclude with Eqs.~(\ref{REK1}) and (\ref{W_N-Z_N})
\begin{eqnarray}
{\cal W}_{GC}^B (\beta,z)  = \gamma_1 (\beta) \,\frac{ \partial }
{ \partial \gamma_1(\beta) }  {\cal Z}_{Gc}^B(\beta,z) \, .
\end{eqnarray}
Taking into account the decomposition (\ref{h_ntren})
we see that according to (\ref{REK5})
\begin{eqnarray}
{\cal W}_{GC}^B (\beta,z) = {\cal Z}_{GC}^B (\beta,z) \sum_{n=1}^{\infty}
\, \gamma_n(\beta) \, z^n  \, .
\label{WGK}
\end{eqnarray}
The canonical ground-state weights can then be found
according to Eq.~(\ref{WGK[z]}) as the
Taylor coefficients following from
\begin{eqnarray}
W_N^B (\beta) \, = \left. \frac{1}{N!} \frac{\partial^N {\cal W}_{GC}^B
(\beta,z)}{\partial z^N} \right|_{z=0} = \,
\sum_{l=1}^N \frac{1}{l! (N\!-\!l)!} \left.  \frac{ \partial^{N-l}
{\cal Z}_{GC}^B (\beta)}{ \partial z^{N-l} } \right|_{z=0} \,
\sum_{n=1}^{\infty} 
\, \gamma_n(\beta) \left. \frac{ \partial^l z^n }{ \partial z^l }
\right|_{z=0} \, .
\label{DDD}
\end{eqnarray}
Using Eqs.~(\ref{REK2}) and
(\ref{REK9}), we obtain the canonical ground-state weights as
\begin{eqnarray}
W_N^B (\beta) = \sum_{n=1}^{N} 
\, \gamma_n (\beta) \, Z_{N-n}^B (\beta) \, ,
\label{W_N}
\end{eqnarray}
which represents the result of Ref.~\cite{Landsberg}.
It states that the weight $W_N^B (\beta)$ of the
particles residing in the ground state of an $N$-particle Bose gas is
immediately known, once the canonical partition functions $Z_n^B (\beta)$ are
calculated for all $n\!<\!N$.
A proper normalization of the statistical ground-state weight
$W_N^B (\beta)$ allows us to determine the probability $w_N^B (\beta)$ to
find a particle in the ground state:
\begin{eqnarray}
w_N^B (\beta) = \frac{ W_N^B (\beta) }{ N Z_N^B (\beta) } \, .
\label{Gr-Warsch}
\end{eqnarray}
This quantity multiplied by $N$ represents
the condensed fraction $N_0$ of an $N$-particle bosonic ensemble.
\section{Special Potentials}\label{POT}
Now we specify the potential energy $V(\bf x)$
and calculate the heat capacity and the ground-state occupancy within canonical
ensembles. In order to see what can go wrong, we first
follow Feynman's textbook on Statistical Mechanics \cite{Feynman} and
describe in Section \ref{HOMOGEN} the homogeneous Bose gas, which leads
to unphysical results. This is corrected
in Section \ref{HOMO-GRUND} by a modified approach for the homogeneous gas
where the special role of the ground state is properly taken into account.
This approach corresponds to the semiclassical treatment of a Bose
gas in a box-like trapping potential, which is then investigated quantum
mechanically exact in Section \ref{BOX}.

Although, the quantities of interest may be calculated in principle
analytically from Eqs.~(\ref{REK12}) and (\ref{W_N}),
the calculations in this section are performed numerically with a C-program
in order to reach large particle numbers. More details concerning analogous
numerical
implementation techniques are presented in Ref.~\cite{Devreese1}.
\subsection{Homogeneous Bose Gas}\label{HOMOGEN}
\begin{figure}[t]
\begin{center}
a)\includegraphics[scale=0.5]{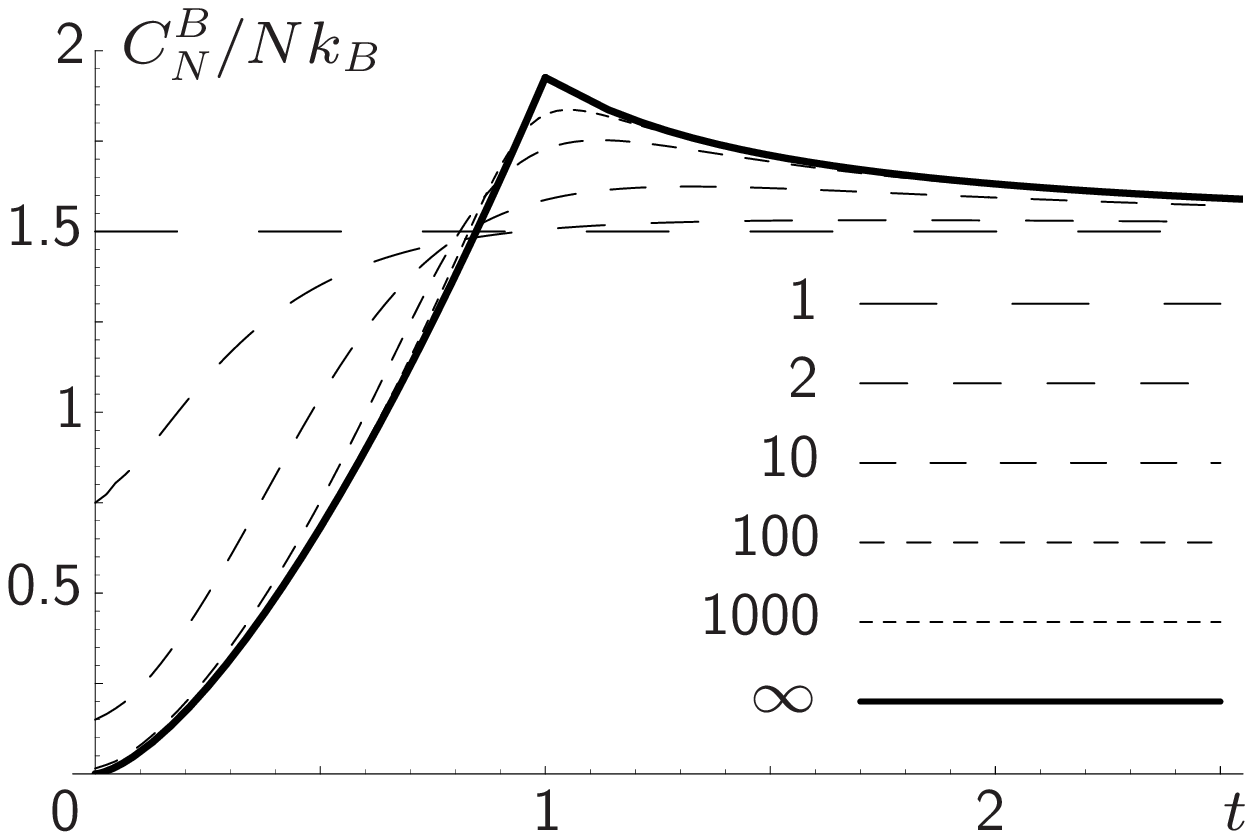}
\hspace*{1.5cm}
b)\includegraphics[scale=0.5]{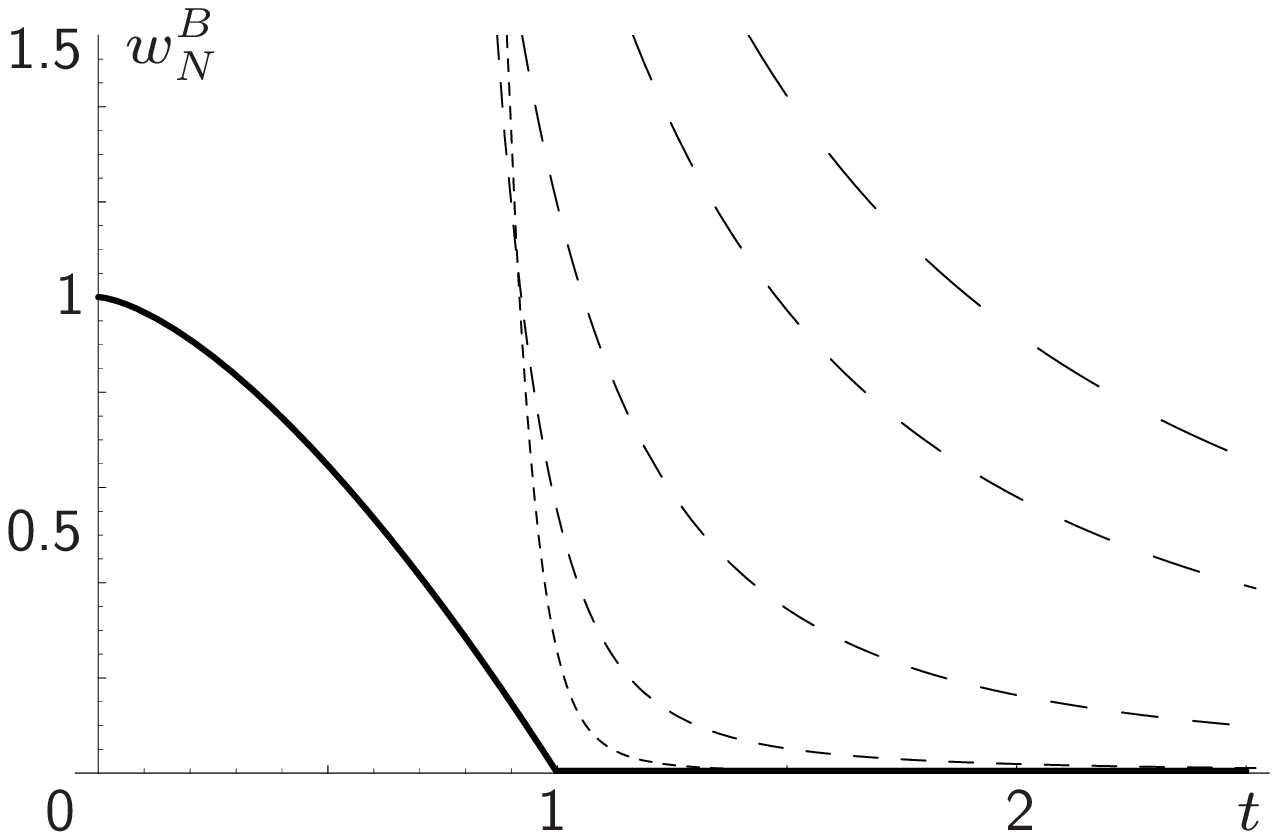}
\end{center}
\caption{\label{Homo-Res}
a) Heat capacity per particle and b) ground-state occupancy versus reduced
temperature in canonical ensembles of $N=$1, 2, 10, 100, 1000 particles
(dashed lines) compared with the grand-canonical result in thermodynamic limit
(solid line). The particle density is chosen to be constant.
}
\end{figure}

Consider  a gas of non-interacting bosons without
any trapping potential, $V( {\bf x} )=0$, i.e. the homogeneous Bose gas.
This theoretical model represents a paradigm studied in the seminal work by
Bose and Einstein \cite{Bose}.
Also the original work of Feynman \cite{Feynman} within the canonical
ensemble relies on treating the homogeneous Bose gas.
But this historical important model has problems, which
manifest themselves in a divergent
ground-state occupancy and in a violation of the third law of
thermodynamics. To exhibit this explicitly, we
insert the free-particle energies
\begin{eqnarray}
E_{\bf k} = \frac{ \hbar^2 }{ 2 M } \, {\bf k}^2 
\label{E-homo}
\end{eqnarray}
with continuous wave vectors ${\bf k}$
into the one-particle partition function (\ref{ZDECO}) and perform
the sum over all energy levels in form of a continuous integral
\begin{eqnarray}
\sum_{\bf k} \to V \int  \, \frac{d^3 k}{(2 \pi)^3} \,\, ,
\label{Sum-Int}
\end{eqnarray}
with $V$ being the volume of the system. The result is
\begin{eqnarray}
Z_1(n \beta) \, = \, \frac{ V }{ \lambda^3 } \, \frac{1}{n^{3/2}} \,\, ,
\label{Z1-homo}
\end{eqnarray}
where $\lambda = \sqrt{ 2 \pi \beta \hbar^2 /M }$ denotes the thermodynamic
de Broglie wave length. According to (\ref{E-homo}) the ground-state energy
vanishes for the homogeneous case, thus leading to the ground-state
contribution (\ref{h_ngrund}) of the partition function
\begin{eqnarray}
\gamma_1(n \beta) = 1 \, ,
\label{gamma-homo}
\end{eqnarray}
which is independent of temperature.
Now we use a dimensionless temperature parameter
\begin{eqnarray}
\tau = V^{2/3} \, \frac{k_B M}{2 \pi \hbar^2} \, T 
\label{tau}
\end{eqnarray}
and represent our results for one and two particles.
From recursion relation (\ref{REK12}) we obtain the canonical
partition functions
\begin{eqnarray}
Z_1^B(\tau) = \tau^{3/2} \,\, , \hspace*{1.2cm}
Z_2^B(\tau) = \, \frac{ \tau^{3/2} }{ 2^{5/2} } + \frac{ \tau^3 }{ 2 } \,\, ,
\label{Z123-homo}
\end{eqnarray}
which also follow from (\ref{QS162}), (\ref{QS16E}) with
(\ref{Z1-homo}). From (\ref{Z123-homo}) we calculate the heat
capacity according to the thermodynamic equation (\ref{C_N}).
Using again the dimensionless temperature parameter
(\ref{tau}) yields for one and two particles
\begin{eqnarray}
C_1^B(\tau) = \, \frac{3}{2} \, k_B  \,\, , \hspace{1.2cm}
C_2^B(\tau) = 3 k_B \, \frac{ 1 + 9 \sqrt{2} \, \tau^{3/2} + 16 \, \tau^3 }
{ 2 + 8 \sqrt{2} \, \tau^{3/2} + 16 \, \tau^3 } \,\, .
\label{C_12}
\end{eqnarray}
From both results we read off that they obey in the high-temperature
limit the classical Dulong-Petit law $C_N^B \to 3 N k_B /2$.
However, in the small-temperature regime one obtains the general result
$C_N^B \to 3 k_B /2$. This finding violates the third law of
thermodynamics, according to which the heat capacity has to vanish at zero
temperature. Both results (\ref{C_12}) are plotted in Figure
\ref{Homo-Res} a) versus the reduced temperature $t \equiv T/T_c$, where 
\begin{eqnarray}
T_c \, = \, \frac{2 \pi \hbar^2}{k_B M} \, \left[ \frac{N}{V \zeta(3/2)}
\right]^{2/3} 
\label{Tc}
\end{eqnarray}
represents the critical temperature of an ideal homogeneous Bose gas.
The reduced temperature $t$ is related to $\tau$ of Eq.~(\ref{tau}) by
\begin{eqnarray}
t \, = \, \left[ \frac{\zeta(3/2)}{N} \right]^{2/3} \tau \, .
\label{t-tau}
\end{eqnarray}
In order to obtain results for larger particle numbers, we evaluate the
recursion relation (\ref{REK12}). Here the variable
(\ref{tau}) is more useful than the reduced temperature (\ref{t-tau}) since
the latter depends explicitly on the particle number. The recursion
(\ref{REK12}) involves results for different particle numbers, so that
the employment of the reduced temperature $t$ would require a permanent
rescaling.
Plotting our results for different particle numbers in Figure
\ref{Homo-Res} a) then only requires a rescaling of the original variable
$\tau$ to the reduced temperature $t$ according
to (\ref{t-tau}).
This plot shows the behavior of the heat capacity per particle for different
total particle numbers. Here, the well-known grand-canonical result
is reached in the thermodynamic limit.
Furthermore, the heat capacity per particle is suppressed at zero temperature
in the thermodynamic limit, but, as mentioned before, the heat capacity
itself remains non-vanishing for $t=0$.

A related but much more dramatic problem arises for the homogeneous Bose gas
in finite systems when we consider the probability to find a particle in the
ground state. To illustrate this we calculate first from (\ref{W_N}) the
ground-state weights
for one and two particles using (\ref{gamma-homo})
\begin{eqnarray}
W_1^B(\tau) = 1 \,\, , \hspace*{1.2cm}
W_2^B(\tau) = 1 + \tau^{3/2} \,\, ,
\label{W123-homo}
\end{eqnarray}
which also follow from (\ref{W_2}) and (\ref{W_3}). The ground-state
probability (\ref{Gr-Warsch}) then yields with (\ref{Z123-homo})
and (\ref{W123-homo})
\begin{eqnarray}
w_1^B(\tau) = \, \frac{1}{\tau^{3/2}}  \,\, , \hspace{1.2cm}
w_2^B(\tau) = \, \frac{2 \sqrt{2}}{\tau^{3/2}} \, \frac{ 1 + \tau^{3/2} }
{ 1 + 2 \sqrt{2} \, \tau^{3/2} } \,\, .
\label{w_12}
\end{eqnarray}
Here we see that the probability to find a particle in the ground state
decays for increasing temperature but diverges for low temperatures.
This behavior is depicted in Figure \ref{Homo-Res} b) where the ground-state
occupancy is also shown
for larger particle numbers. One can also see that our canonical results
do not match the well-known grand-canonical shapes in the thermodynamic limit.
%
%
%
%
Moreover, it diverges for all finite systems at zero temperature.
Not only that this divergency is physically meaningless, but by definition
any probabilities have to be bounded by unity. In the present description of
the homogeneous Bose gas in the canonical ensemble this is obviously not the
case. As it will be seen in the next subsection, this problem appears
due to an improper treatment of the semiclassic limit and the results can be
corrected
by taking care of the
special role of the ground state.
\subsection{Homogeneous Bose Gas with Ground State} \label{HOMO-GRUND}
\begin{figure}[t]
\begin{center}
a)\includegraphics[scale=0.5]{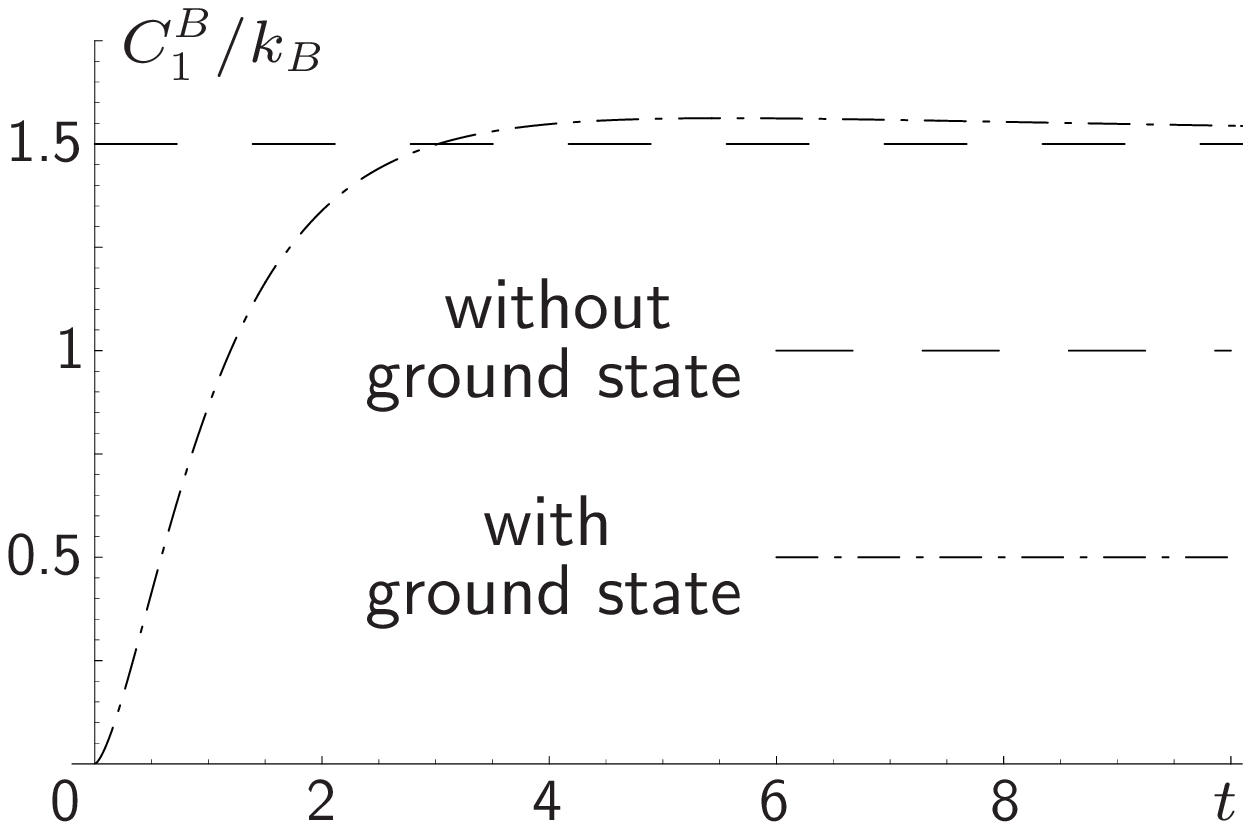}
\hspace*{1.5cm}
b)\includegraphics[scale=0.5]{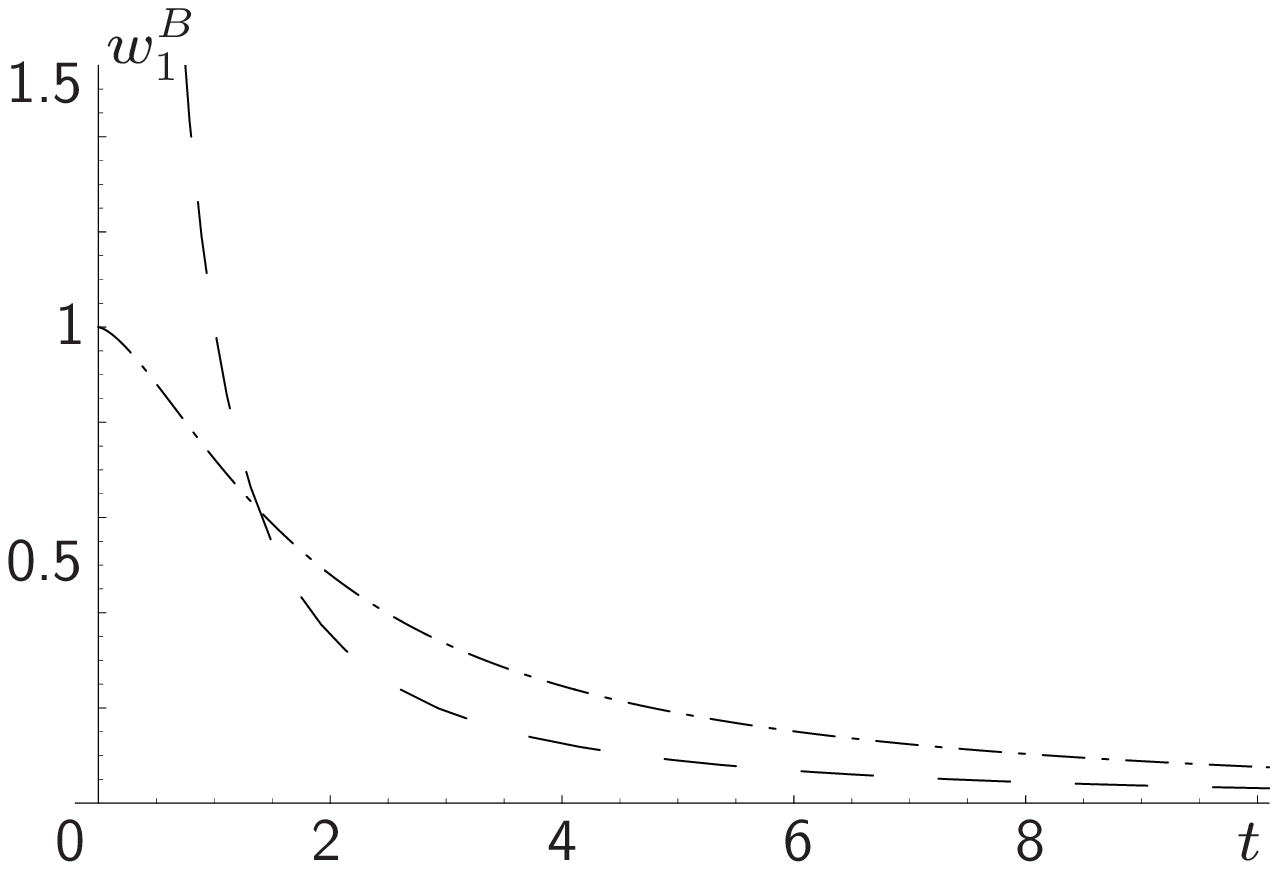}
\end{center}
\caption{\label{Homo-Homogrund}
a) Heat capacity and b) ground-state probability for one particle versus
reduced temperature calculated
without (dashed line) and with an extra ground state (dashed-dotted line).
}
\end{figure}

%
%
Inspecting the results in the previous subsection more carefully, we
recognize
already at the beginning a contradiction between the one-particle partition
function (\ref{Z1-homo}) and its ground-state contribution
(\ref{gamma-homo}). Namely, whereas (\ref{Z1-homo})
vanishes for low temperatures, the ground-state partition remains finite.
This leads to a negative contribution for all excited states. However, one
would reasonably expect that the latter has to be always positive and
vanishes only for zero temperature. According to this expectation we suppose
that the partition function has to be a sum of the ground-state contribution
(\ref{gamma-homo}) and the contribution of excited states in the form given by
(\ref{Z1-homo}):
\begin{eqnarray}
Z_1(n \beta) \, = \, 1 + \, \frac{ V }{ \lambda^3 } \, \frac{1}{n^{3/2}} \,\, .
\label{Z1-grund}
\end{eqnarray}
Such a decomposition is equivalent to replace condition
(\ref{Sum-Int}) by 
\begin{eqnarray}
\sum_{\bf k} f({\bf k}) \, \to \, f({\bf 0}) +
V \! \int  \, \frac{d^3 k}{(2 \pi)^3} \, f({\bf k}) 
\label{Sum-new}
\end{eqnarray}
for any function $f({\bf k})$. Comparing (\ref{Sum-new}) with relation
(\ref{Sum-Int}) we see that the ground state in the latter didn't play
any special role. There we integrated over all states including the ground state,
so that the latter was only a point of zero measure, as basically any other
state. In our present approach, however, we separate the ground state explicitly from
the remaining excited states.
In order to demonstrate how well this procedure works, we calculate
the heat capacity (\ref{C_N}) of one particle from (\ref{Z1-grund}), yielding
for $D=3$ spatial dimensions
\begin{eqnarray}
C_1^B(\tau) = \, \frac{3 k_B}{2} \, \frac{ 5 \, \tau^{3/2} + 2 \, \tau^3 }
{ 2 + 4 \, \tau^{3/2} + 2 \, \tau^3 }  \,\,  .
\label{C_1-grund}
\end{eqnarray}
This leads to $3 k_B/2$ in the high-temperature limit according to the
Dulong-Petit law. For small temperatures the heat capacity (\ref{C_1-grund})
vanishes satisfying also the third law of thermodynamics. Both results
(\ref{C_12}) and (\ref{C_1-grund}) for the heat capacity of one particle are
compared in Figure \ref{Homo-Homogrund} a).

The second problem of the homogeneous Bose gas in the last subsection
concerns the probability to find a particle in the ground state which
diverges for vanishing temperature.
Taking into account the extra ground state, we recognize that (\ref{gamma-homo})
represents its contribution to the partition function (\ref{Z1-grund}).
Thus, the probability for a particle to be in the ground state
seems to be normalized. For one particle we use
Eq.~(\ref{W_N}) to obtain $W_1^B(\tau) = \gamma_1(\tau)$, yielding with
the definition (\ref{Gr-Warsch})
the ground-state probability
\begin{eqnarray}
w_1^B(\tau) = \, \frac{1 }{ 1 + \tau^{3/2}}  \,\, .
\label{w_1-grund}
\end{eqnarray}
It vanishes for high temperature ($\tau \gg 1$) and has its maximal value
of one for zero temperature, as one may expect on general physical grounds.
We compare this result with the corresponding one of the previous subsection
in Figure \ref{Homo-Homogrund} b). For small temperatures the corrected curve
does not diverge and shows the expected behavior.
One understands immediately, why this procedure works so
well for finite systems. Assuming the particle density to be equal for all
systems, the volume $V$ has to be small for small particle numbers. The
additional unity in (\ref{Z1-grund}) plays for them a more important role
than for larger ensembles. Moreover, this unity is practically negligible for
high enough temperatures and carries a larger weight for lower temperatures.
As the contributions of different lower particle numbers are mixed in the
quantities of interest (\ref{REK12}) and (\ref{W_N}), this
influences the situation decisively.

\begin{figure}[t]
\begin{center}
a)\includegraphics[scale=0.5]{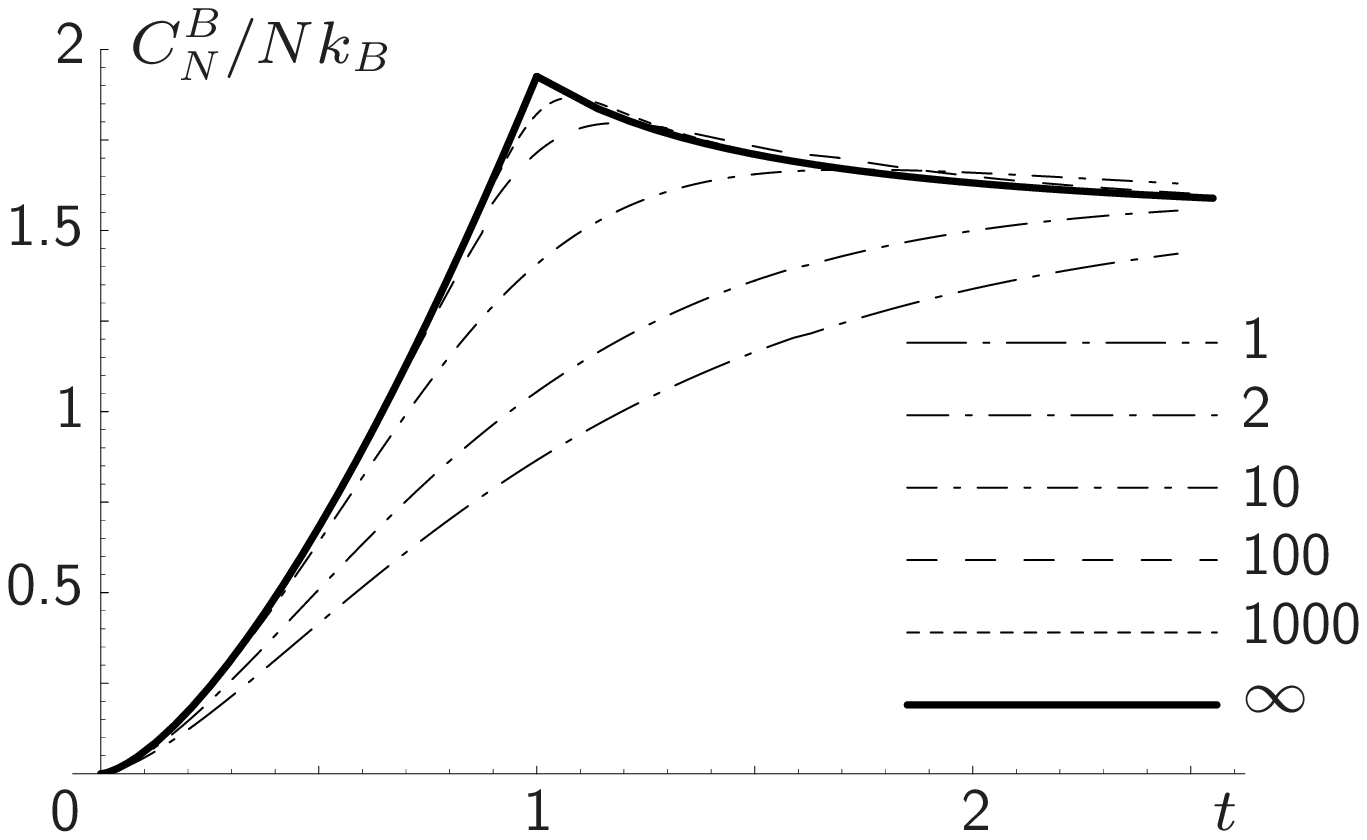}
\hspace*{1.1cm}
b)\includegraphics[scale=0.5]{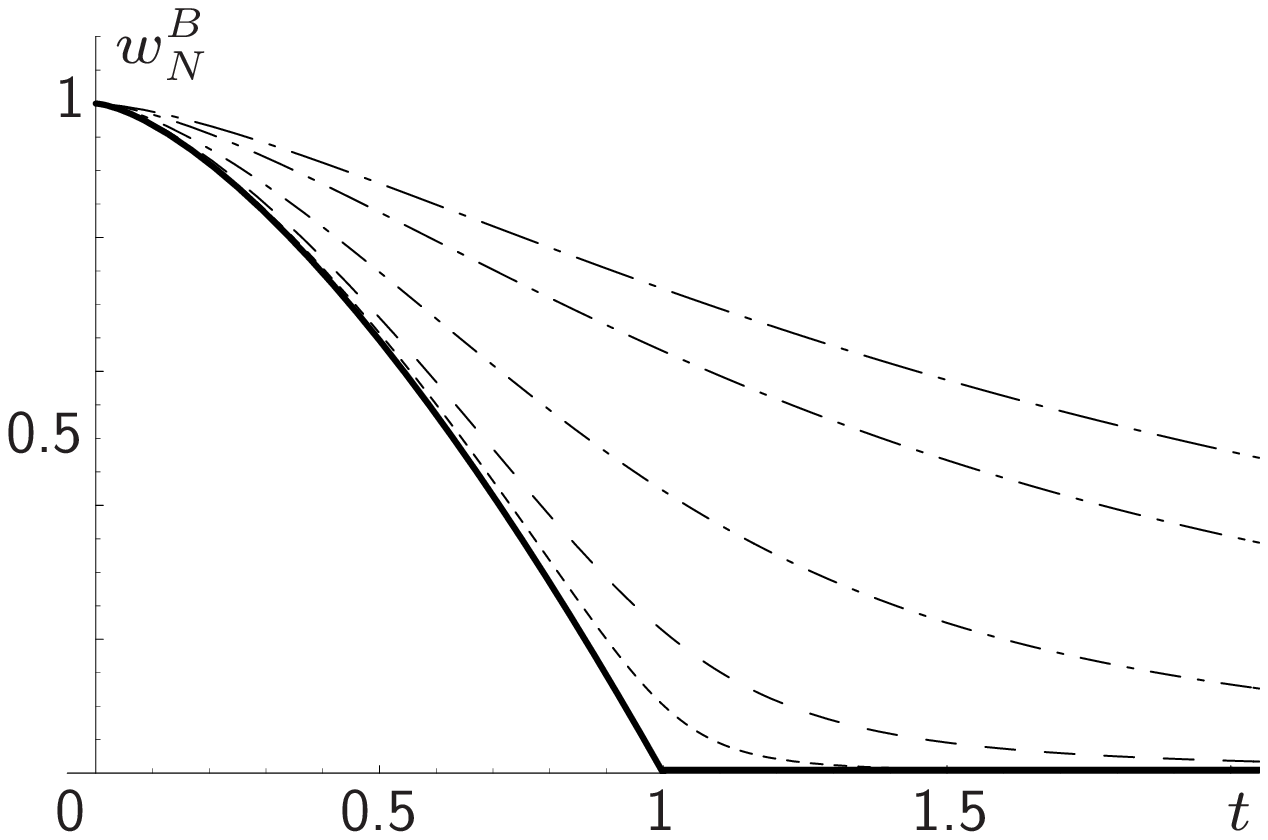}
\end{center}
\caption{\label{Grund-Res}
a) Heat capacity per particle and b) ground-state occupancy
versus reduced temperature calculated with an
extra ground state according to (\ref{Z1-grund}). Results are represented for
canonical ensembles of $N=1,2,10,100,1000$ particles (dashed and dashed-dotted
curves) and for the grand-canonical ensemble in thermodynamic limit
(solid curve) for a fixed particle density.
}
\end{figure}

Now we turn to many-body systems and calculate their heat capacity according
to the recursion relation (\ref{REK12})
with definition (\ref{C_N}) and the ground-state probability according
to (\ref{W_N}) with (\ref{Gr-Warsch}). The corresponding results are
represented in Figure \ref{Grund-Res}. All curves have the correct behavior
for high temperatures as well as for the low-temperature region. Moreover, in
the thermodynamic limit, where the particle number goes to infinity,
both results display the grand-canonical shape. This means that, although
all curves are smooth for finite systems, the behavior of the heat capacity
and the ground-state occupancy change more pronounced at the critical point
by increasing the number of particles.

Thus, we can summarize that a careful treatment of the ground state in
this subsection improves the description of the homogeneous Bose gas in a fundamental way.
Nevertheless, also this attempt represents only a physical approximation to a
real finite system. To this end we will show in the next subsection that the
exact quantum mechanical results deviate quantitatively in a significant way
from the semiclassical treatment presented here.
\subsection{Bose Gas in Box Potential}\label{BOX}
In this section we work out the exact quantum mechanical treatment of a
Bose gas in a square-box potential. For convenience, we choose the size of
our box to be equal to $L$ in all spatial directions, so that the trapping
potential reads
\begin{eqnarray}
  V({\bf x}) \, = \, \left\{ \begin{array}{ll}
  \hspace{2mm} 0 \hspace{4mm} ,  & 0 \leq  x_j  \leq L \hspace{3mm}
  {\rm for \,  all} \hspace{2mm} j=1,2,3  \, ,
\\*[2mm]
  \,\, \infty \hspace{3mm} , & \hspace{11mm}
  {\rm elsewhere}  \, .
\end{array} \right.
\label{VKast}
\end{eqnarray}
The one-particle Schr{\"o}dinger equation 
is then solved exactly
by the energy eigenvalues
\begin{eqnarray}
E_{\bf k} = \, \frac{ \pi^2 \hbar^2 }{ 2 \, M L^2 } \, {\bf k}^2 \hspace*{5mm}
{\rm with} \hspace*{4mm} k_j = 1,2,3,4,... \hspace*{3mm} (j=1,2,3) \,\, .
\label{E-Kast}
\end{eqnarray}
As one can deduce from the normalization condition of the eigenstates, the
value ${\bf k} = (k_1,k_2,k_3)$ is not a solution, if at least one $k_j$
vanishes. Thus, the ground-state energy is given by
$E_G = 3 \pi^2 \hbar^2 /( 2 M \! L^2 )$. From this we obtain the
ground-state contribution to the partition function (\ref{h_ngrund})
which reads with the dimensionless temperature parameter (\ref{tau}) as
\begin{eqnarray}
\gamma_1(n \beta) = e^{ - 3 n \pi /(4 \tau) } \, .
\label{gamma-box}
\end{eqnarray}
The one-particle partition function then reads
\begin{eqnarray}
Z_1(n \beta) \, = \left[ \, \sum_{m=1}^{\infty} e^{ - n \pi m^2 /(4 \tau)}
\right]^3  \, .
\label{Z1-box}
\end{eqnarray}
The corresponding quantities in the Subsections \ref{HOMOGEN} and
\ref{HOMO-GRUND} are different semiclassical approximations of
(\ref{gamma-box}) and (\ref{Z1-box}), which are justified for an infinitely
large box. Such a semiclassical approximation leads, at first,
to the ground-state contribution (\ref{gamma-homo}) and, secondly, to a
vanishing level spacing which gives rise to the continuously distributed energy levels
(\ref{E-homo}). Therefore, the sum in (\ref{Z1-box}) has to
be replaced by an integral according to (\ref{Sum-Int}) or (\ref{Sum-new}),
respectively. The one-particle partition function is then given by a Gaussian
integral and can be simply calculated leading to (\ref{Z1-homo}) or
(\ref{Z1-grund}). Strictly speaking, this semiclassical description of
a finite system is inadequate insofar as it is impossible to have a finite
number of particles in an infinite box, unless in the trivial case of zero
density.

\begin{figure}[t]
\begin{center}
\includegraphics[scale=0.45]{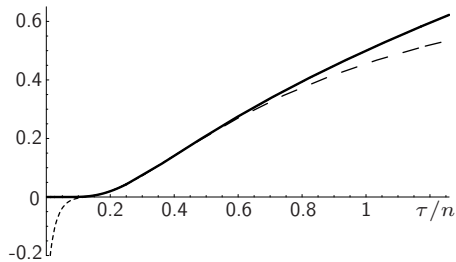}
\end{center}
\caption{\label{Schnitt}
Comparison between the exact sum
$\sum_{m=1}^{\infty} e^{- n \pi m^2/ (4 \tau)}$
(solid line) and their lowest small-temperature approximation
$e^{- n \pi / (4 \tau)}$ (long dashed curve) according to (\ref{Z1-box}) and
the lowest high-temperature approximation
$\sqrt{\tau /n} - 1/2 + 2 \sqrt{\tau /n} \, e^{- 4 \pi \tau /n}$
(short dashed curve) from (\ref{Z1-dual}).
}
\end{figure}

The partition function (\ref{Z1-box}) in a box cannot be calculated exactly.
Thus, we can evaluate the sum only approximatively up to a certain order of
the summation index $m$. However, this procedure does not work well
for all temperatures since the exponential functions
$e^{- n \pi /(4 \tau)}$ in (\ref{Z1-box}) are required to be small, and this
is only true
for small temperatures ($\tau \ll 1$). In order to perform the sum
in (\ref{Z1-box}) also for high temperatures, we need its dual expression,
which follows from the Poisson summation formula \cite{Kleinert1}
\begin{eqnarray}
\sum_{m=-\infty}^{\infty} f(m) \, =  \sum_{q=-\infty}^{\infty}
\int_{-\infty}^{\infty} d z \, f(z) \, e^{2 \pi i q z} \, .
\label{Poisson}
\end{eqnarray}
With this,
the one-particle partition function (\ref{Z1-box}) is converted into
\begin{eqnarray}
Z_1(n \beta) = \left[ \, \sqrt{ \frac{\tau}{n} } \, - \, \frac{1}{2} + 2 \,
\sqrt{ \frac{\tau}{n} } \sum_{q=1}^{\infty} \, e^{ - 4 \pi \tau q^2 /n }
\right]^3  \, .
\label{Z1-dual}
\end{eqnarray}
Of course, this function cannot be calculated exactly
either, so we have to perform
the summation also here only up to a certain order. Whereas the
approximation (\ref{Z1-box}) is applicable for small temperatures, its dual
(\ref{Z1-dual}) can be evaluated for high temperatures.
How can we now use both representations (\ref{Z1-box}) and (\ref{Z1-dual})
in order to evaluate the one-particle partition function in the whole
temperature range? In principle, both representations are equivalent, but
only when both series are evaluated exactly.
This exact expression is approximated well by summing few terms in
(\ref{Z1-box}) only below a certain temperature and in (\ref{Z1-dual})
only above a certain temperature. There exists an appropriate
temperature region where
both approximations coincide. In Figure \ref{Schnitt} we see that
for the interval $0.1< \tau/n <0.5$ even the lowest term in (\ref{Z1-box})
as well as (\ref{Z1-dual}) are sufficient.
By taking more terms into account, this matching region
grows. For practical calculations, however, the summation of many terms costs much effort, so
it is more efficient to precisely define an
intermediate
temperature, which joins the low-temperature region with the high-temperature
region best.

\begin{figure}[t]
\begin{center}
a)\includegraphics[scale=0.5]{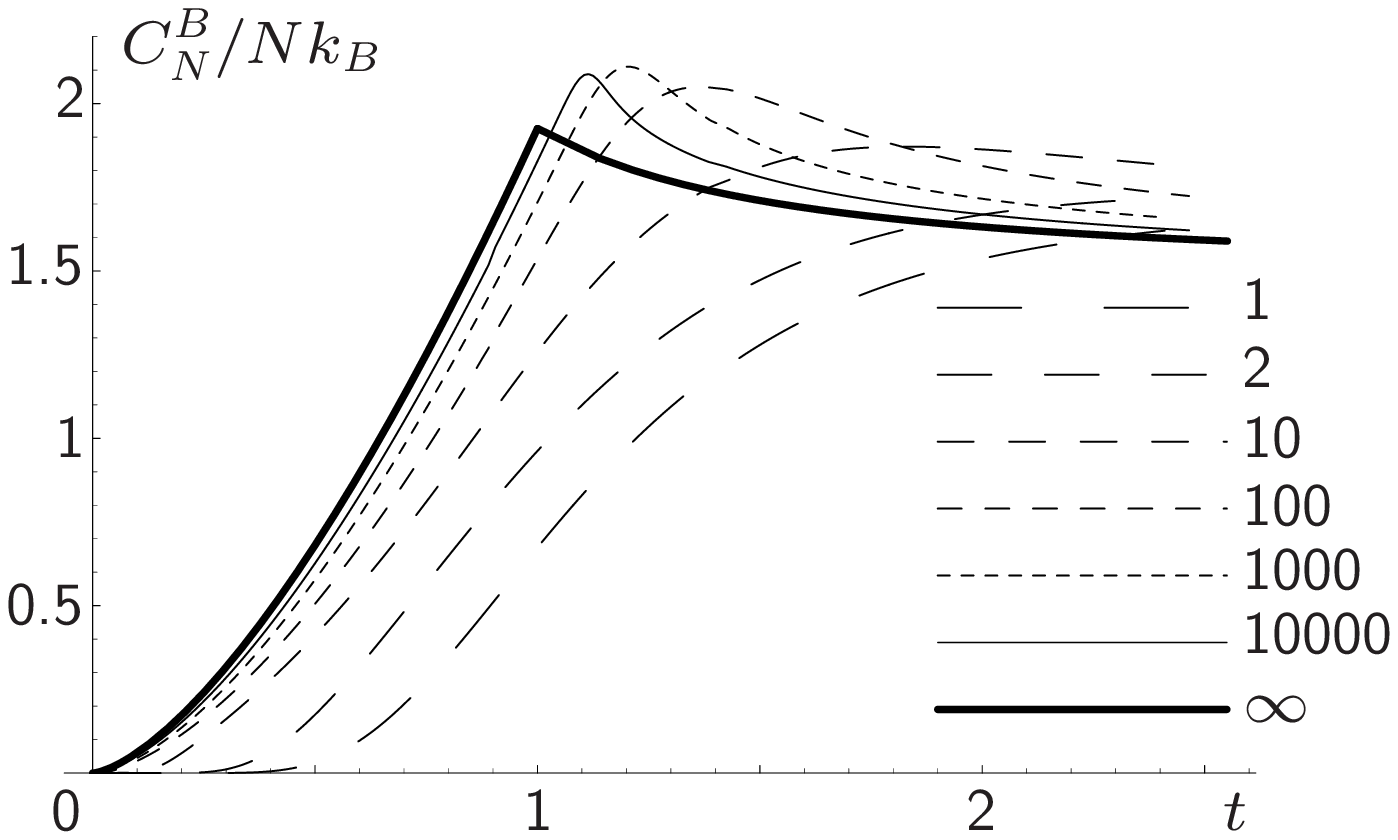}
\hspace*{1.1cm}
b)\includegraphics[scale=0.5]{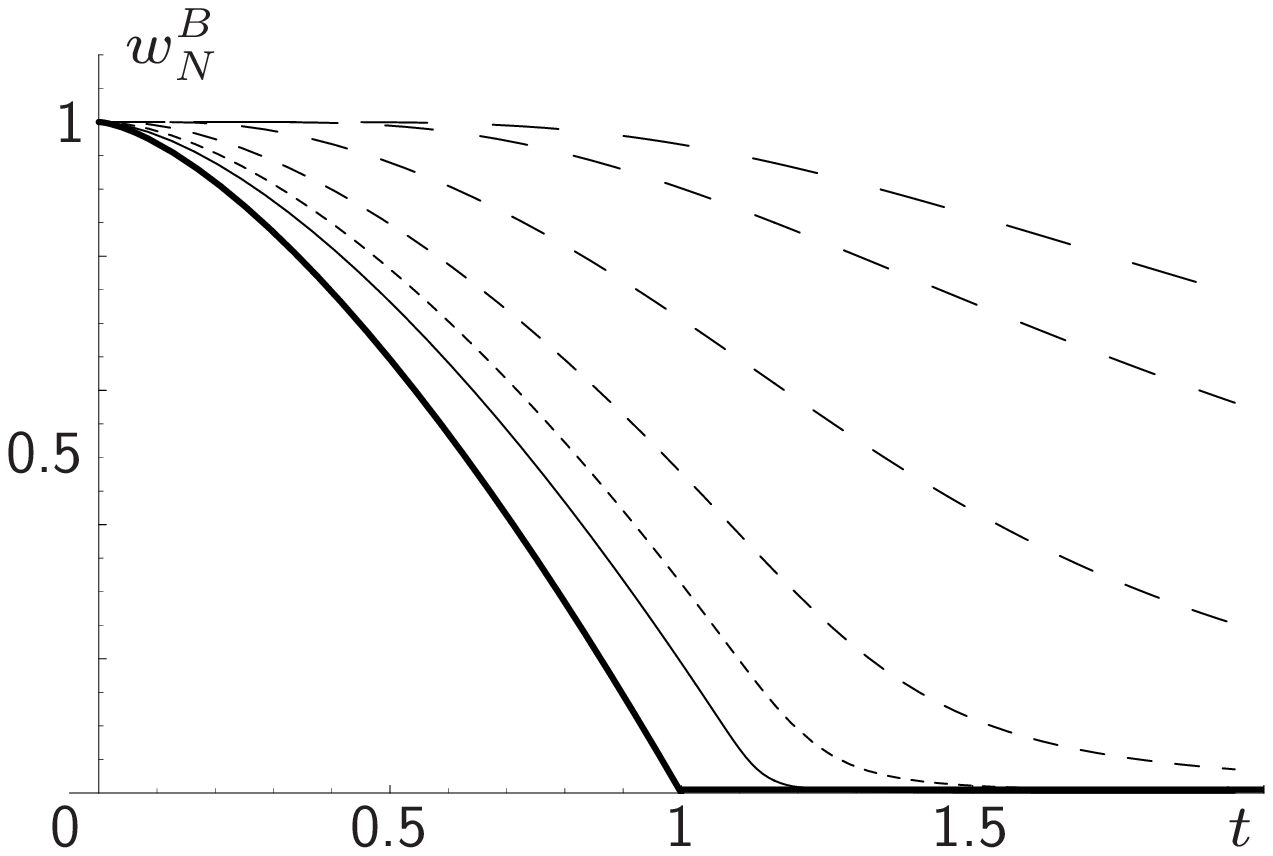}
\end{center}
\caption{\label{Box-Res}
a) Heat capacity per particle and b) ground-state occupancy
versus reduced temperature in exact square-box
potential for ensembles of $N=1,2,10,100,1000,10000$ bosons (dashed and thin
curves) and in the grand-canonical ensemble in thermodynamic limit (thick
solid curve). All results are calculated for the same particle density.
}
\end{figure}

In order to calculate the results for many particles we use the recursion
relation (\ref{REK12}) for the partition
functions and (\ref{W_N}) for the ground-state weights. The heat
capacity is then calculated from (\ref{C_N}) and the ground-state occupancy
from (\ref{Gr-Warsch}). Results for $N=1,2,10,100,1000$ and $10000$ bosons are
depicted in Figure
\ref{Box-Res}. All results for the heat capacity exhibit the Dulong-Petit
behavior in the high-temperature limit and fulfill the third law of
thermodynamics for $T \to 0$. The ground-state occupancy is $100 \%$
for zero temperature and vanishes for high temperatures. One
specific feature of an exact quantum mechanical treatment is worth emphasizing.
Due to the non-vanishing gap between the ground state and the first excited state,
the slope of the heat capacity is exponential for small temperatures.
This is especially visible for small particle numbers $N$. For higher $N$,
the exponential region decreases and is followed by the usual power-law
increase $\sim t^{3/2}$ . In the thermodynamic limit the exponential region
disappears and the power behavior takes over.

As expected, the canonical results converge in the thermodynamic limit towards
the grand-canonical one for an infinite average particle number. However, this
convergence is much slower in the
quantum-mechanical exact treatment than in the semiclassical approximation
of the previous subsection.
This manifests itself, for instance, in an apparent
discrepancy in the location and extent of the heat capacity maximum even
for $N=10000$ particles. The same observation holds for the
ground-state occupancy. Although an increase of the particle number makes the
curves sharper in the crossover region, the location of
the latter is clearly shifted towards higher temperatures. This gives
rise to the so-called finite-size effect which will be elaborated in the
subsequent section.
\section{Finite-Size Effect}\label{Tkrit}
\begin{figure}[t]
\begin{center}
\includegraphics[scale=0.55]{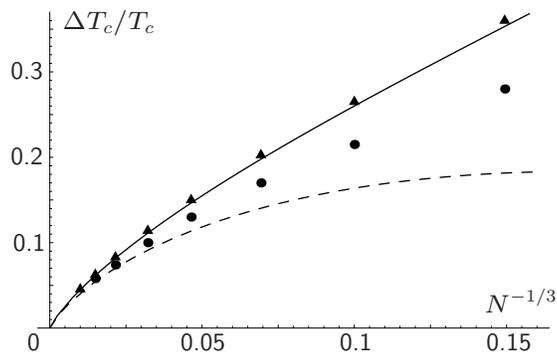}
\end{center}
\caption{\label{Semi-Res}
Shifts of critical temperature for $N$ bosons in
square-box potential with $N=300$, $1000$, $3000$, $10000$, $30000$, $100000$,
$300000$ (from right to left) in canonical ensemble (dots).
They differ considerably from the grand-canonical results (triangles) with
the same boson numbers. The dashed curve corresponds to the semiclassical
result (\ref{SemiHolt}) within grand-canonical ensemble from
Ref.~\cite{Holthaus}. The solid curve represents our result
$\Delta T_c/T_c = t_c - 1$ according to Eq.~(\ref{N-t}).
}
\end{figure}

Let us now investigate in detail the
crossover temperature, which can be regarded as a generalized critical
temperature $T_c$. At first, we suggest to define this as the
temperature where the ground-state occupancy possesses the largest
curvature. This definition is motivated by the fact that
in the thermodynamic limit this quantity has an
infinite curvature at the critical point. By analogy, in a finite
ensemble the point with a significantly sharp rise in the curvature may play
the role of the transition
temperature. These points are plotted with dots in Figure \ref{Semi-Res} for
different particle numbers.
The finite-size effects are quite large.
In a box confinement, they amount to $30 \%$ for $N=300$ and to about $5 \%$
for $N=300000$ bosons. The effects here are about
3-4 times larger than for the Bose gas confined in a harmonic trap
\cite{Grossman}. Furthermore, it is remarkable that
the finite-size shifts of $T_c$ point upwards which is opposite to the harmonic case.
This fact can be
explained by a stability analysis, which shows that in a 3D-box potential
only finite systems are stable \cite{Yukalov}. Particle fluctuations
grow with the size of the box, resulting in a loss of stability, and this
is why bigger systems condense
at lower temperatures.

Let us compare our canonical results for the critical temperature with
analytic approximations. The first estimate of the shift of the critical
temperature
was given in Ref.~\cite{Holthaus} by
\begin{eqnarray}
\frac{\Delta T_c}{T_c} \, = \, \frac{ 1 }{ \zeta^{2/3}(3/2) \, N^{1/3} } \,
\ln \left[ \frac{4 \, N^{2/3}}{3 \pi \zeta^{2/3}(3/2)} \right] + \, \dots
\,\, .
\label{SemiHolt}
\end{eqnarray}
This result is plotted
in Figure \ref{Semi-Res} as a dashed curve.
It represents the grand-canonical result for semiclassical corrections caused
by large but finite particle numbers $N$,
which takes into account the non-vanishing energy eigenvalues $k_j$ of
(\ref{E-Kast}) in a box potential. This correction was calculated in
Ref.~\cite{Holthaus} by replacing the
sum over all energy eigenstates in (\ref{MT6B}) by a momentum integration
and subtracting all
states lying at coordinate planes in the momentum space.
As one can see in Fig. \ref{Semi-Res}, this analytic result reproduces
well the direction of the critical temperature shift but
drastically underestimates its magnitude.
Further corrections could be obtained by avoiding
double-countings at cross sections of coordinate planes as well as
by taking into account the finite gap between the ground state and the first
excited states.

Here we prefer a more controllable procedure for calculating the finite-size
corrections which goes as follows.
We start with the cycle number representation of the grand-canonical average
of the particle number according to Eqs.~(\ref{REK5}) and (\ref{MT6B}):
\begin{eqnarray}
\label{NGK}
\langle N \rangle = \sum_{n=1}^{\infty} Z_1(n \beta) z^n = \,
\langle N_0 \rangle + \langle N_T \rangle \,\, .
\end{eqnarray}
It contains the number of particles in the ground-state condensate
\begin{eqnarray}
\label{N0}
\langle N_0 \rangle = \sum_{n=1}^{\infty} e^{- n \beta E_G} z^n \, = \,
\frac{ 1 }{ z^{-1} e^{\beta E_G} - 1 }
\end{eqnarray}
and the number of thermal particles in excited states
\begin{eqnarray}
\label{Nexc}
\langle N_T \rangle = \sum_{n=1}^{\infty} \left[ Z_1(n \beta) -
e^{- n \beta E_G} \right] z^n \, .
\end{eqnarray}
Furthermore, we rewrite the one-particle partition function
(\ref{Z1-box}) as
\begin{eqnarray}
Z_1(n \beta) \, = e^{ - 3 n \pi /(4 \tau)} \left[ 1 + e^{ - 3 n \pi /(4 \tau)}
\sigma(\tau/n) \right]^3  
\label{Z1-sigma}
\end{eqnarray}
with dimensionless temperature (\ref{tau}) and an excitation contribution
\begin{eqnarray}
\sigma(\tau) = \sum_{m=2}^{\infty} e^{ - \pi (m^2 -4) /(4 \tau)} = \,
\sqrt{\tau} - \, 3/2 \,+ \pi/\sqrt{\tau} \,+\, {\cal O}(1/\tau)
\,\, ,
\label{sigma}
\end{eqnarray}
where the latter equation is the high-temperature approximation similar to
the duality relation (\ref{Z1-dual}) for the whole partition function. Since
we aim at finding the location of the critical temperature, we demand for
the fugacity $z$ the condition
$z e^{- \beta E_G} = z e^{ - 3 \pi /(4 \tau)} \approx 1$
but at the same time $\langle N_0 \rangle \ll \langle N \rangle$. Both
conditions are exactly fulfilled in the thermodynamic limit and are reasonable
approximations for large but finite systems. From
Eqs.~(\ref{NGK})--(\ref{sigma}) we
obtain at the critical point $\tau=\tau_c$ approximately
\begin{eqnarray}
\langle N \rangle \, = \, \sum_{n=1}^{\infty} \left[ \, 3 \,
e^{ - 3 n \pi/(4 \tau_c)} \sigma(\tau_c/n) + 3 \, e^{ - 6 n \pi/(4 \tau_c)}
\sigma^2(\tau_c/n) + e^{ - 9 n \pi/(4 \tau_c)} \sigma^3(\tau_c/n) \, \right]
\,\, .
\label{Nkrit}
\end{eqnarray}
According to Eq.~(\ref{sigma}) the sums in (\ref{Nkrit}) can be expanded in
terms of the polylogarithmic Bose functions $\zeta_{\nu}(e^{-x}) \equiv
\sum_{n=1}^{\infty} e^{-n x} /n^{\nu} $ as
\begin{eqnarray}
\langle N \rangle \!&=&\! \tau_c^{3/2} \, \zeta_{3/2} \left(
e^{- 9 \pi/(4 \tau_c)} \right) + \, \frac{3 \, \tau_c}{2} \left[ \, 2 \,
\zeta_{1} \left( e^{- 6 \pi/(4 \tau_c)} \right) - 3 \,
\zeta_{1} \left( e^{- 9 \pi/(4 \tau_c)} \right) \, \right]
\no \\*[2mm]
&&\!\!+ \, \frac{3 \, \tau_c^{1/2}}{4} \left[ \, 4 \,
\zeta_{1/2} \left( e^{- 3 \pi/(4 \tau_c)} \right) - 12 \,
\zeta_{1/2} \left( e^{- 6 \pi/(4 \tau_c)} \right) + (4 \pi + 9) \,
\zeta_{1/2} \left( e^{- 9 \pi/(4 \tau_c)} \right) \, \right] +
\,\, ...
\label{Nkrit-zeta}
\end{eqnarray}
For large values of $\tau_c$, the polylogarithmic functions are evaluated
with the help of the Robinson formula \cite{Robinson}
\begin{eqnarray}
\zeta_{\nu}(e^{-x}) \, = \,
\left\{ \begin{array}{ll}
{\displaystyle \Gamma(1-\nu) \, x^{\nu - 1} + \, \sum_{k=0}^{\infty} \,
(-x)^k \, \zeta( \nu - k ) /k!} \,\,\,
& {\rm for} \hspace{2mm} 1 \neq \nu < 2  \, ,
\\*[3mm]
{\displaystyle - \ln x + \, \sum_{k=1}^{\infty} \, (-x)^k \, \zeta( 1 - k )
/k!} \,\,\,
& {\rm for} \hspace{5mm} \nu = 1  \, ,
\end{array} \right.
\label{ROB}
\end{eqnarray}
where $\Gamma(z)$ denotes the Gamma function and
$\zeta(z) \equiv \sum_{n=1}^{\infty} \, 1/ n^{z}$ the Riemann zeta
function.
Inserting this into (\ref{Nkrit-zeta}) and identifying
$\langle N \rangle$ with the actual particle number $N$, we arrive at
\begin{eqnarray}
N \, = \, \tau_c^{3/2} \, \zeta(3/2) + \frac{3}{2} \, \tau_c
\, \ln \left[ C_3 \, \frac{\pi}{2 \tau_c} \right] + \, \frac{3(\pi \!+\! 1)}{4}
\, \tau_c^{1/2} \, \zeta(1/2) \, + \,
{\cal O}(\tau_c^{0}) \,\, .
\label{N-tau}
\end{eqnarray}
Using the reduced temperature $t \equiv T/T_c$ instead of $\tau$ according to
(\ref{t-tau}), this can also be rewritten as
\begin{eqnarray}
t_c^{3/2} \, = \, 1 + \frac{3}{2} \, \frac{ t_c }{ N^{1/3} \zeta^{2/3}(3/2) }
\, \ln \left[ \frac{ 2 N^{2/3} }{\pi \zeta^{2/3}(3/2) } \, \frac{t_c}{C_3}
\right] - \, \frac{3(\pi \!+\! 1)}{4} \, \frac{ t_c^{1/2} \, \zeta(1/2) }
{ N^{2/3} \, \zeta^{1/3}(3/2) }
\, + \, {\cal O}(1/N) \,\, .
\label{N-t}
\label{@RESU}\end{eqnarray}
Taking into account only the first term on the right-hand side yields the
semiclassical result $t_c=1$. The first correction
involves a coefficient $C_3$ which unfortunately cannot be calculated
directly. The origin of this problem is the divergency of the
polylogarithmic function $\zeta_{\nu}(z)$ for $z \approx 1$ and any
$\nu \leq 1$, which is explicitly given by the first term in (\ref{ROB}).
So one can see that the last three terms in Eq. (\ref{Nkrit-zeta}) contribute
not only to the order of $\tau_c^{1/2}$ but also to the order $\tau_c$. The
same is valid for all other terms which we omitted in (\ref{Nkrit-zeta}). The
leading finite-size correction of $\tau_c$ plays here a special role
insofar as it consists of infinitely many terms and thus its factor $C_3$
cannot be calculated from the first few terms. However, we can perform a direct numerical evaluation,
yielding the value $C_3 \approx 0.9574$ which also follows from a semi-analytic
calculation \cite{Kleinert1}. In contrast to that, all higher corrections in
(\ref{N-t}) are analytically calculable.

In fact, such a problem is not typical for the box potential only.
In a harmonic trap, one encounters an analogous situation for the subleading
finite-size correction instead of the leading one for the present box case.
Here, we should mention, that it is just this non-analyticity of the leading
finite-size effect, which is
responsible for the failure of the approximation in
Ref.~\cite{Holthaus}, see Eq.~(\ref{SemiHolt}) here. Their perturbative 
approach  yielded the factor
$3/2$ instead of the numerical correct $C_3$-value.

By extracting the $N$-dependence of the reduced critical temperature
of Eq.~(\ref{N-t}), we now obtain a grand-canonical semiclassical result
which is represented by the solid line in Figure \ref{Semi-Res}.
As one can see, it overestimates canonical results considerably.
In fact, although many physical properties within canonical and
grand-canonical ensemble are equal in the thermodynamic limit, they
can be quite different for finite systems. Let us now determine
crossover temperatures for finite grand-canonical ensembles
in the same way as before in the canonical approach.
To this end we calculate the ground-state occupancy for a grand-canonical
system. As already outlined at the beginning of Section \ref{KONDENS}, this 
is done by determining at first the fugacity $z$ which corresponds to a 
given average particle
number $\langle N \rangle$ for a fixed temperature
from (\ref{NGK}) and inserting its value into (\ref{N0}). The resulting
quantity $\langle N_0 \rangle /\langle N \rangle$ is a smooth function
of temperature similar to the canonical results in Fig. \ref{Box-Res} b)
but somewhat shifted to the right.
The crossover temperatures are then calculated by analogy with the
canonical case as points of largest curvature and are indicated by
solid triangles in Figure
\ref{Semi-Res}. They agree very well with the analytic approximation of
(\ref{N-t}) even for smaller systems of a few hundred particles,
but differ strongly from canonical values.

It is well-known that the canonical partition function can be obtained
from grand-canonical partition function by means of the
integral over the chemical potential $\mu$
using
the identity
%
%
\begin{eqnarray}
Z_N^B = \beta \int_{c-i\infty}^{c+i\infty}\frac{d\mu}{2\pi i}\,
e^{- \beta \mu N}
{\cal Z}_{GC}^B(\mu)
\,\, .
\label{Resid}
\end{eqnarray}
In the leading saddle-point or steepest-descent approximation one obtains
\begin{eqnarray}
Z_N^B = e^{- \beta \bar\mu N}
{ {\cal Z}_{GC}^B(\bar\mu) }
+ \, ... \,\, ,
\label{sattel}
\end{eqnarray}
where $\bar\mu$ is the extremal point of the exponent
$-\beta \mu N+\log Z _{GC}^B(\mu)$ in (\ref{Resid}).
The extrimity condition implies by means of Eq.~(\ref{ZGK}) immediately 
the usual Bose-Einstein distribution (\ref{MT6B}).
For infinite $N$, this approximation becomes exact,
leading to the equality of thermodynamical properties \cite{Ziff,Huang}
in canonical and grand-canonical ensembles.

For finite $N$, one may want to calculate
corrections from the neighborhood of the saddle point.
The naively perturbation up to first order, howewer, yields a
wrong sign of the shift for the critical temperature. A detailed
analysis shows that a non-analyticity occurs at the saddle point near
the critical point
and infinitely many orders in the saddle point expansion
have to be calculated and resummed to find the correct result.
This fact has been discussed in literature, for instance in
\cite{Ziff} for the box potential and in \cite{Dingle} for a general case
near the critical point, and is not yet solved. 
\section{Conclusion}
By taking into account the
special role of the ground state of the system, we have obtained a consistent
semiclassical description of an ideal Bose gas of
a fixed finite number of particles
confined in a box. We have thus improved
Feynman's original canonical approach
for treating the ideal homogeneous Bose gas.
Finite-size effects have been shown to be significant,
even for a moderate number of particles. Our
canonical findings are compared with grand-canonical
ones, for which we have found analytic expressions
which improve results in the literature. An analogous analytic expression
within the canonical ensemble could not be found, so we have to be content 
with numerical results, which were given in the present paper.

Our canonical
investigation may serve as a theoretical basis for future experiments
in box-like traps manufactured with strongly inhomogeneous fields or
laser beams.
\section*{Acknowledgment}
We thank the DFG
Priority Program SPP 1116 {\it Interactions in Ultra-Cold Atomic and
Molecular Gases} for financial support.

\end{document}